\documentclass[10pt,journal,letterpaper]{IEEEtran}
\IEEEoverridecommandlockouts
\usepackage[T1]{fontenc}
\usepackage{cite}
\usepackage{amsmath,amssymb,amsfonts}
\usepackage{algorithmic}
\usepackage{graphicx}
\usepackage[colorlinks,citecolor=blue,linkcolor=blue,urlcolor=blue]{hyperref}
\usepackage{xcolor}
\usepackage{orcidlink}
\usepackage{setspace}

\definecolor{blue}{rgb}{0.1, 0.2, 0.92}
\definecolor{green}{rgb}{0.13,0.55,0.13}
\definecolor{red}{rgb}{1,0,0}
\begin{document}

\bstctlcite{MyBSTcontrol}

\title{Stokes Parameters and Dual Classical-Quantum Signaling}%
\author{Anjali~Dhiman,~\IEEEmembership{Student~Member,~IEEE}, Ziqing~Wang,~\IEEEmembership{Member,~IEEE}, Timothy~C.~Ralph, \\Ryan~Aguinaldo,~\IEEEmembership{Member,~IEEE}, and
Robert~Malaney,~\IEEEmembership{Senior~Member,~IEEE}\\%
\thanks{The Australian Government supported this research
through the Australian Research Council Linkage Projects
funding scheme (Project No. LP200100601). The views expressed here are those of the authors and are not necessarily those of the Australian Government or the Australian Research Council.  A.D. is supported by the Sydney Quantum Academy. Approved for Public Release: Distribution is unlimited; \#25-0913

Anjali Dhiman, Ziqing Wang, and Robert Malaney are with the School of Electrical Engineering and Telecommunications, University of New South Wales, Sydney, NSW 2052, Australia. Timothy C. Ralph is with the Centre for Quantum Computation and Communication Technology, School of Mathematics and Physics, University of Queensland, St Lucia, QLD 4072, Australia. Ryan Aguinaldo is with Northrop Grumman Corporation, San Diego, CA 92128, USA.}}
\maketitle

\begin{abstract} 
Catering to emerging satellite-based free-space optical (FSO) communication networks and exploiting polarization encoding via Stokes operators, we propose a novel simultaneous quantum-classical communications (SQCC) protocol. The protocol enables the coexistence of secure quantum communications and high-throughput classical communications with minimal alterations in both the infrastructure and the energy input. Compared to the conventional SQCC protocol, our new approach provides superior practicality in the real world, eliminates the need for a separate local oscillator, and allows for the simple readout of both quantum and classical information using direct detection. The protocol also minimizes the undesirable interplay between the quantum and the classical parts of communication. We provide a detailed mathematical formulation of the protocol, along with theoretical and numerical analysis of its performance,  illustrating a promising path to practical and effective realization of combined classical-quantum communications.{\footnote{This paper is an extended version  of \cite{GCversion}.}}
\end{abstract} 

\section{Introduction}
Quantum communications represent a unique paradigm for information security due to their underlying dependence on quantum mechanics. 
As the first proposed application of quantum communications, the Quantum Key Distribution (QKD)~\cite{R1} can offer unconditional information security that classical protocols cannot achieve~\cite{Shor2000,Gisin2002}. 
In recent years, extensive research efforts have focused on satellite-based QKD schemes as a means of extending quantum communications to a global scale ~\cite{Bedington2017,nauerth2013air,Liao2017,Yin2017,Dequal2020}. 
As an alternative to discrete-variable QKD (which encodes quantum information into discrete features of single photons, see, e.g.,~\cite{R1,ZiqingTMQKD,ZiqingOAMQKD,Tao2021MitigatingTE,flamini2018,Kish2020}), continuous-variable (CV) QKD (which encodes quantum information onto the optical-field quadrature variables)~\cite{Laudenbach2017,R2,R3,Jaksch2024,Hajomer2024Experiment,Neda2021} offers a practical advantage since CV quantum states can be readily generated and measured with high-speed and high-efficiency optical devices.
    
The CV paradigm in the optical domain has further enabled the concept of simultaneous quantum-classical communications (SQCC)~\cite{Qi2016,Qi2018}, where quantum (typically CV-QKD) and classical communications can be implemented using the same communication infrastructure and the same optical pulse with minimum additional complexity (see proof-of-principle experiments in,~e.g.,~\cite{SQCC_25kmFiber}).
Despite its promising potential, the originally proposed implementation of SQCC~\cite{Qi2016,Qi2018} (hereafter referred to as \textit{conventional} SQCC) is not immune to issues. Most notably, conventional SQCC requires a separate local oscillator (LO) to measure the quadratures at the receiver.\footnote{In conventional SQCC, the LO can be multiplexed with the signal at the transmitter (see e.g.,~\cite{SQCC_25kmFiber}). It is also possible to construct different measurement schemes that reconstruct the LO locally at the receiver (see, e.g.,~\cite{Qi2018}), thus avoiding some known attacks on the LO (see, e.g.,~\cite{LO_Loophole1,LO_Loophole2}). Similar locally constructed schemes could be built around our new protocol, but we do not consider them here.} This LO generally traverses the system hardware in a different spatial path from the signal, leading to likely phase mismatches detrimental to CV-QKD.  The conventional SQCC is also subject to an interaction (crosstalk) between its quantum and classical parts, resulting in a complicated performance trade-off between them~\cite{Nick2025}.  In this work, we explore a novel protocol that implements SQCC without such issues

Unlike a fiber channel, the Earth atmosphere is non-birefringent and preserves the polarization state of light to a large extent. 
Although less explored, polarization (which has been widely used to encode DV quantum information in the single-photon regime) can also be used for CV quantum encoding via the modulation of quantum Stokes operators~\cite{Vidiella2006}. When used to describe the quantum polarization properties of intense light fields, these operators can be treated as field quadratures~\cite{R23,schnabel2003stokes}. 
In addition to providing inherent robustness against most of the deteriorating effects imposed by the turbulent atmosphere of the Earth, when engineered appropriately, 
CV polarization encoding via Stokes operators sends the LO co-propagating with the quantum signal through polarization multiplexing. This approach allows for simple readout of quantum information based on direct detection via standard photodiodes~\cite{R23}.
The feasibility of CV-QKD encoded by polarization has been confirmed by experimental demonstrations (see, e.g.~\cite{Lorenz2004,Lorenz2006,Elser2009,heim2010,Stokes2014Discrete,Shen2019,Zheng2023}) with a real-world demonstration over an FSO channel of $460\,\text{m}$~\cite{Shen2019}. 

Despite providing many unique advantages in FSO CV-QKD, the use of polarization encoding via Stokes operators has never been explored for the purpose of satellite-based SQCC -- in this work, we do just that. The novel contributions of this work are as follows. 
\begin{enumerate}
 \item We introduce a new protocol that uses polarization encoding via Stokes operators for both the classical and quantum encoding (hereafter referred to as the \textit{Stokes-based SQCC} protocol). Our new design offers superior real-world practicability by (i) eliminating the need for sending a separate LO, (ii) allowing for the simultaneous readout of both the quantum and the classical information using simple direct detections, and (iii) minimizing the undesirable interplay between the quantum and classical parts of communications.
\item We detail the operation procedures and propose a practical implementation of our protocol.
In order to achieve maximum practicability, we map each step of our protocol to a feasible physical implementation, provide the corresponding mathematical description, and explain important practical implications where necessary.
 \item We perform a detailed performance investigation on our new protocol over FSO channels, confirming its usefulness and superiority via both theoretical analysis and numerical validation in both the asymptotic and composable finite-size regimes.
\end{enumerate}

The remainder of this paper is as follows. Section~\ref{Sec:Background} introduces the Stokes operators and our classical-quantum encoding scheme. Section~\ref{Sec:SystemModel} describes the system model of our Stokes--based SQCC protocol. Section~\ref{Sec: MathForm} describes the mathematical framework of our system model. Section~\ref{Sec:Asymp_Finite-sizeKeys} provides a detailed mathematical description of the secret key rate in the asymptotic and finite-size regime. Section~\ref{Sec:Results} presents the numerical results on the performance of our protocol over FSO channels. We conclude in Section~\ref{Sec:Conclusions}. 

\section{Background}\label{Sec:Background}
Using the $H$ (horizontal) and the $V$ (vertical) directions as a basis, a transverse electromagnetic field can arbitrarily be decomposed into two components corresponding to orthogonal polarization modes. We consider three common decompositions that describe the same transverse electromagnetic field with $H$ and $V$ components, diagonal ($D$) and anti-diagonal ($A$) components, and right-handed circular ($R$) and left-handed circular ($L$) components. As a result, the polarization state of a beam of light can be described using the four Stokes parameters, namely ${S}_0$ (the total intensity of the $H$ and $V$ components), ${S}_1$ (the intensity difference between the $H$ and $V$ components), ${S}_2$ (the intensity difference between the $D$ and $A$ components), and ${S}_3$ (the intensity difference between the $L$ and $R$ components).   In quantum optics, after changing intensities to photon-number operators, the corresponding Stokes operators are given by~\cite{R23}
\begin{equation}\label{Eq:S0}
\scalebox{0.9}{%
   $\hat{S}_0= \hat{a}^\dagger_H \hat{a}_H+\hat{a}^\dagger_V \hat{a}_V=\hat n_H+ \hat n_V,$}
\end{equation}
\begin{equation}\label{Eq:S1}
\scalebox{0.9}{%
   $\hat{S}_1= \hat{a}^\dagger_H \hat{a}_H-\hat{a}^\dagger_V \hat{a}_V= \hat n_H- \hat n_V,$}
\end{equation}
\begin{equation}\label{Eq:S2}
\scalebox{0.9}{%
   $\hat{S}_2= \hat{a}^\dagger_H \hat{a}_V +\hat{a}^\dagger_V \hat{a}_H= \hat n_D- \hat n_A,$}
\end{equation}
\begin{equation}\label{Eq:S3}
\scalebox{0.9}{%
   $\hat{S}_3= i(\hat{a}^\dagger_V \hat{a}_H -\hat{a}^\dagger_H \hat{a}_V) =\hat n_R- \hat n_L,$}
\end{equation}
where $\hat{a}$ ($\hat{a}^\dagger$) denotes the annihilation (creation) operator, and $\hat{n}$ denotes the photon-number operator ($\hbar =2$ henceforth). Note that, in Eqs.~(\ref{Eq:S0}-\ref{Eq:S3}), the subscripts of the operators denote the above-defined polarization components. The Stokes operator $\hat{S}_0$ commutes with the other three Stokes operators, that is, $[\hat{S}_0,\hat{S}_i]=0$ ($i=1,2,3$). The commutator of the remaining three Stokes operators is given by
$[\hat{S}_j,\hat{S}_k]=2i\epsilon_{jkl}\hat{S}_l\,(j, k, l= 1,2,3)$, leading to the corresponding uncertainty relations $\operatorname{Var}(\hat{S}_j)\operatorname{Var}(\hat{S}_k) \geq |\epsilon_{jkl}\langle\hat{S}_l\rangle|^2$ (with $\operatorname{Var}(\hat{S}_j)=\langle \hat{S}_j^2\rangle-\langle\hat{S}_j\rangle^2$ being the variance of each Stokes operator). This dictates that it is not possible to simultaneously measure any two of the Stokes operators with certainty as long as the third one is nonzero.
The four Stokes operators form the  Stokes vector,  \textbf{S}, given by ($\hat {S}_0$,$\hat {S}_1$,$\hat {S}_2$,${\hat S}_3$)$^T$.

 In Eqs.~(\ref{Eq:S0}-\ref{Eq:S3}), the annihilation and creation operators can be expressed as $\hat{a}_{M}=\alpha_{M}+\delta\hat{a}_{M}$ ($M \in\{H,V\}$) where $\alpha_{M} = \langle \hat a_M \rangle$ 
is the classical amplitude and $\delta\hat a_{M}$ represents the quantum fluctuation of the $M$ component with $[\delta\hat{a}_{M},\delta\hat a_{M}^\dagger]=1$ and $\langle\delta\hat{a}_{M} \rangle=0$. 
Thus, Eqs.~(\ref{Eq:S0}-\ref{Eq:S3}) can be re-expressed as
\begin{equation}\label{Eq:S0_1}
\scalebox{0.9}{%
$\hat S_0 =|\alpha_H|^2+|\alpha_V|^2+\alpha_H\delta{\hat{X}}^+_H +\alpha_V\delta{\hat{X}}^+_V,$} 
\end{equation}
\begin{equation}\label{Eq:S1_1}
\scalebox{0.9}{%
$\hat S_1 =|\alpha_H|^2-|\alpha_V|^2+\alpha_H\delta{\hat{X}}^+_H -\alpha_V\delta{\hat{X}}^+_V, $}
\end{equation}
\begin{equation}\label{Eq:S2_1}
\scalebox{0.9}{%
   $\hat S_2 =2\alpha_H\alpha_V+\alpha_H\delta{\hat{X}}^+_V +\alpha_V\delta{\hat{X}}^+_H,$}
\end{equation}
\begin{equation} \label{Eq:S3_1}
\scalebox{0.9}{%
   $\hat S_3 =i(2\alpha_H\alpha_V)+\alpha_V\delta{\hat{X}}^-_H +\alpha_H\delta{\hat{X}}^-_V,$}
\end{equation}               
where{\footnote{In Eq.~(\ref{Eq:S3_1}), a typographical error in \cite{GCversion} is corrected.} $\delta \hat{X}_{M}^+=\delta \hat{a}_{M}^\dagger + \delta \hat{a}_{M}$, $\delta \hat{X}_{M}^{-}=i(\delta \hat{a}_{M}^\dagger - \delta\hat{a}_{M})$, and $\langle \delta \hat{X}_{M}^+\rangle = \langle \delta \hat{X}_{M}^-\rangle=0$. Note that in deriving Eqs.~(\ref{Eq:S0_1}-\ref{Eq:S3_1}) we have assumed that $\alpha_{M}$ is a real number with $|\alpha_{M}|\gg 1$ and therefore we only considered the first-order fluctuation terms\footnote{Second-order fluctuations terms are discussed in the  Appendix~A.}. 
The variance of the Stokes operators is then given by
\begin{equation}
\scalebox{0.8}{%
$\operatorname{Var}(\hat{S}_0)=\alpha_H^2 \langle(\delta{\hat{X}}_H^+)^2\rangle +  \alpha_V^2 \langle(\delta{\hat{X}}_V^+)^2\rangle + 2\alpha_H \alpha_V \langle\delta{\hat{X}}_H^+\delta{\hat{X}}_V^+\rangle,$}
\end{equation}
\begin{equation}\label{Eq.VS1}
\scalebox{0.8}{%
$\operatorname{Var}(\hat{S}_1)=\alpha_H^2 \langle(\delta{\hat{X}}_H^+)^2\rangle + \alpha_V^2 \langle(\delta{\hat{X}}_V^+)^2\rangle -   2\alpha_H \alpha_V \langle\delta{\hat{X}}_H^+\delta{\hat{X}}_V^+\rangle,$}
\end{equation}
\begin{equation}\label{Eq.VS2}
\scalebox{0.8}{%
$\operatorname{Var}(\hat{S}_2)=\alpha_H^2 \langle(\delta{\hat{X}}_V^{+})^2\rangle +\alpha_V^2 \langle(\delta{\hat{X}}_H^{+})^2\rangle   
    + 2\alpha_H \alpha_V \langle\delta{\hat{X}}_V^{+}\delta{\hat{X}}_H^{+}\rangle,$}
\end{equation}
\begin{equation}\label{Eq.VS3}
\scalebox{0.8}{%
$\operatorname{Var}(\hat{S}_3) =\alpha_H^2 \langle(\delta{\hat{X}}_V^{-})^2\rangle +\alpha_V^2 \langle(\delta{\hat{X}}_H^{-})^2\rangle   
    + 2\alpha_H \alpha_V \langle\delta{\hat{X}}_V^{-}\delta{\hat{X}}_H^{-}\rangle.$}
\end{equation}
Here, we explain the theory of using the Stokes operators for our SQCC encoding, which is based on the use of a strongly polarized optical beam. 
Specifically, for classical encoding, we modulate the Stokes operator $\hat{S}_1$ by mapping a binary classical bit to the dominating direction of polarization (either $H$ or $V$) of an optical beam -- we encode bit `1' (`0') by setting $|\alpha_H|^2 \gg |\alpha_V|^2\approx 0$ ($|\alpha_V|^2 \gg |\alpha_H|^2\approx 0$), leading to $\langle \hat{S}_1\rangle\!\approx\!|\alpha_H|^2$ ($\langle \hat{S}_1\rangle\!\approx\!-|\alpha_V|^2$) and $\langle \hat{S}_1\rangle\!\approx\!\langle\hat S_0\rangle$. For QKD encoding, we modulate the normalized Stokes operators $\hat{S}_2^{\prime}= {\hat{S}_2}/{\langle\hat{S}_1\rangle}^{1/2}$ and $\hat{S}_3^{\prime}= {\hat{S}_3}/{\langle\hat{S}_1\rangle}^{1/2}$ -- these two operators satisfy $[\hat S_2^{\prime},\hat S_3^{\prime}]=2i$ and thus resemble the conjugate field quadratures (i.e., $\hat{Q}$ and $\hat{P}$) used for quantum encoding in coherent-state CV-QKD~\cite{R2}.

\section{System Model}\label{Sec:SystemModel}
As introduced in Section~\ref{Sec:Background}, the main theme of our Stokes-based SQCC protocol is to use $\hat S_1$ as a carrier of classical information and $\hat S_2, \hat S_3$ for the purpose of quantum communications (QKD in particular). 
\begin{figure*}[h!]
\centering
\includegraphics[width= 1.0\linewidth]{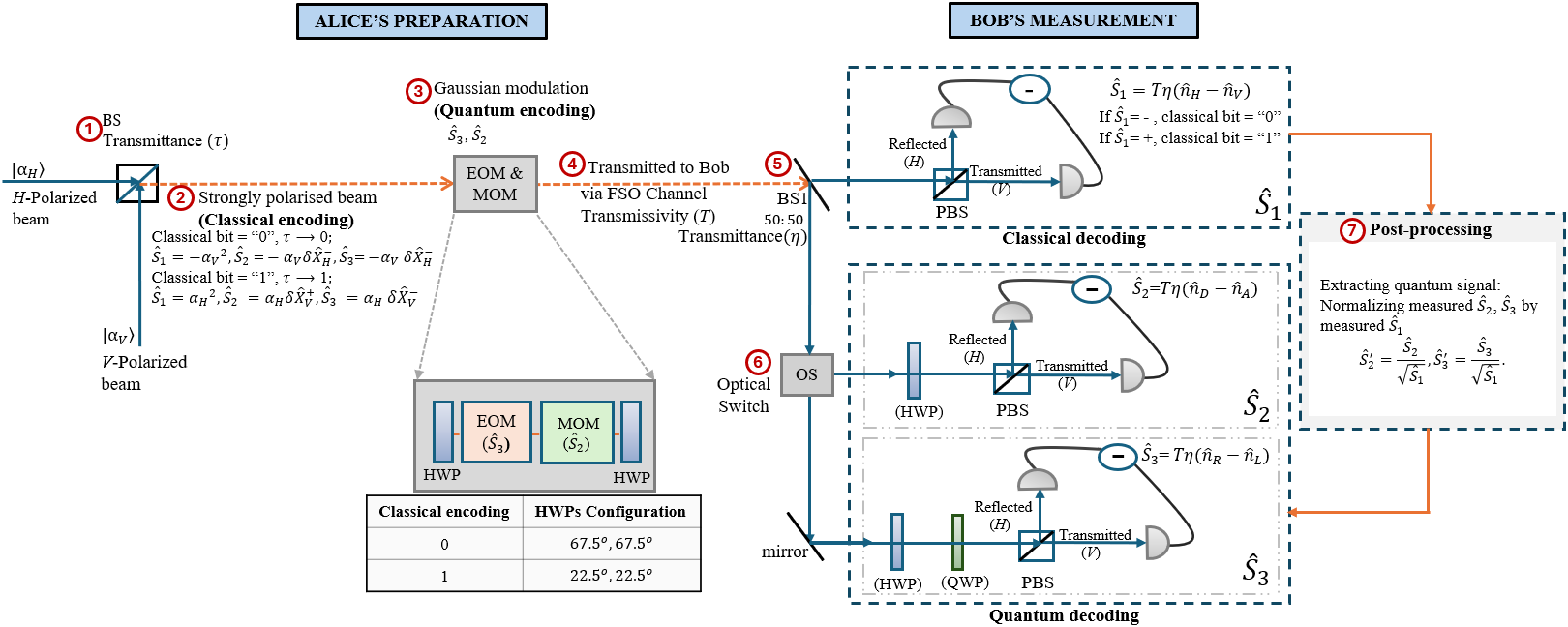}
    \caption{Schematic of our {{Stokes--based SQCC protocol}}.
    {\raisebox{.5pt}{\textcircled{\raisebox{-.9pt} {1}}}} $H$ and $V$-polarized beams prepared and directed to a BS with variable transmittance $\tau$.
    {\raisebox{.5pt}{\textcircled{\raisebox{-.9pt} {2}}}} If the encoded classical bit is ``0'' (``1''), $\tau$ approaches 0 (1), and the emerging beam is highly polarized in $V$ ($H$)-direction.
    {\raisebox{.5pt}{\textcircled{\raisebox{-.9pt} {3}}}}  Beam passes through the HWPs, EOM, and MOM for the modulation of $\hat S_3$ and $\hat S_2$, respectively, as shown in the zoomed-in Grey box. Both HWPs are set at $67.5^o$ if Alice encodes classical bit ``0'' and at $22.5^o$ if she encodes classical bit ``1''.  {\raisebox{.5pt}{\textcircled{\raisebox{-.9pt} {4}}}} Beam is sent to Bob through an FSO channel with transmissivity $T$. 
    {\raisebox{.5pt}{\textcircled{\raisebox{-.9pt} {5}}}} Received beam passes through BS1 (transmittance $\eta$) to be split into two. 
    The classical message is decoded by passing the transmitted component of BS1 through a PBS, followed by direct detection.
    {\raisebox{.5pt}{\textcircled{\raisebox{-.9pt} {6}}}} The reflected component of BS1 is used for quantum decoding. An optical switch is used to randomly switch between the measurement of $\hat S_2$ and the measurement of $\hat S_3$ (equivalent to homodyne detection). $\hat S_2$ is measured using a HWP and a PBS followed by direct detection. $\hat S_3$ is measured using a HWP, a QWP, and a PBS followed by direct detection.  
    {\raisebox{.5pt}{\textcircled{\raisebox{-.9pt} {7}}}} The measured value of $\hat S_2$ and $\hat S_3$ in step {\raisebox{.5pt}{\textcircled{\raisebox{-.9pt} {6}}}} is normalized using the measured value of $\hat S_1$ in step {\raisebox{.5pt}{\textcircled{\raisebox{-.9pt} {5}}}.}}
   
\label{Fig.Schematic}
\end{figure*}

In this section, we detail the implementation setup and operational procedures of our protocol. Section ~\ref{Sec: InitialStatePrep} to \ref{Sec:QuantDeco} describes the first method of implementation, while Section ~\ref{Sec: InitialStatePrep2} describes the second method. A schematic describing the implementation of our protocol using the first method is presented in Fig.~\ref{Fig.Schematic} and Fig.~\ref{Fig:AliceState} represents the second. Our analysis will assume an effective monochromatic picture, even though we envisage most CV-QKD systems to be based on pulsed laser systems (delivering key rates in bits/pulse). We can reconcile these two systems by assuming the pulses to be of long duration in the sense that a pseudo-monochromatic approximation holds. Clearly, for broadband pulses (narrow temporal profiles) such as emerging frequency-comb systems, this approximation will not be valid. In the following, we will refer to the encoding of a pulse as a ``shot.''

\subsection{ State preparation via controlled combining} \label{Sec: InitialStatePrep}
We first introduce a state preparation scheme based on the controlled combination of two orthogonally polarized beams for classical encoding.
Our protocol starts with the initial preparation of the unmodulated state at Alice.
For each shot of classical-quantum modulation, Alice prepares two classical beams, one polarized in the $H$-direction and the other in the $V$-direction, with the same classical amplitude $\alpha$. We assume that there is some practical energy constraint that limits $\alpha$ (so classical communication is not completely error-free; see later discussion). Alice ensures that her two beams are distinguishable only in their polarization and directs them to a beam splitter (BS) with variable transmittance $\tau$ as shown in Step {\raisebox{.5pt}{\textcircled{\raisebox{-.9pt} {1}}}} of Fig.~\ref{Fig.Schematic}. 

\subsection{Classical encoding} \label{Sec: ClassEnc}
We first briefly describe the corresponding classical encoding and quantum encoding schemes supported by the aforementioned intuitive initial state preparation method. Specifically, 
Alice first performs classical encoding by properly controlling the combining of the two prepared polarized beams at the BS. 
This corresponds to Step {\raisebox{.5pt}{\textcircled{\raisebox{-.9pt} {2}}}} in Fig.~\ref{Fig.Schematic}. The controlled combining can be achieved by a wavelength-independent tunable directional coupler (e.g.,~\cite{ArbitaryCoupler}) -- in a practical implementation, such a coupler can be replaced by a simpler setup where an electronically controlled variable optical attenuator is used to control each of the inputs of a standard beam splitter.
Specifically, to encode the classical bit ``0'' (``1''), Alice adjusts $\tau$ to $\tau \sim 0$ ($\tau \sim 1$), making the output beam (emerging from the transmission port of the BS) strongly polarized in $V$-direction ($H$-direction) and weakly polarized in $H$-direction ($V$-direction). At this stage, the output beam state can be considered as a coherent two-mode state. 

Consider a scenario in which Alice encodes the classical bit ``1'' by setting $\tau \sim 1$, leading to $|\alpha_H|^2 = |\alpha|^2 \gg |\alpha_V|^2\approx 0$. 
The Stokes operators in Eqs.~(\ref{Eq:S0_1}-\ref{Eq:S3_1}) then become
\begin{gather}
\hat S_0 \approx  \hat S_1 =|\alpha_H|^2+\alpha_H\delta{\hat{X}}^+_H,  \label{Eq:S0_AC}\\
   \hat S_2=\alpha_H\delta{\hat{X}}^+_V,\,\,\,\hat S_3  =\alpha_H\delta{\hat{X}}^-_V.\label{Eq:S3_AC}
\end{gather}
Since the beam is strongly polarized along the $H$-direction, the second term of Eq.~(\ref{Eq:S0_AC}) can be neglected. 
After further normalization of $\hat S_2$ and $\hat S_3$, Eqs.~(\ref{Eq:S0_AC}-\ref{Eq:S3_AC}) become
\begin{equation}
    \hat S_1 =|\alpha_H|^2,\label{Eq:S1_AC} 
\end{equation}
\begin{equation}
   \hat S_2^{\prime} =\delta{\hat{X}}^+_V=\delta{\hat{X}}^+_{V, \text{vac}},\label{Eq:S2_normalised}
\end{equation}
\begin{equation}
   \hat S_3^{\prime}=\delta{\hat{X}}^-_V=\delta{\hat{X}}^-_{V, \text{vac}}\label{Eq:S3_normalised}.
\end{equation}
Similarly, if Alice intends to encode ``0'', she will set $\tau \sim 0$ to give $|\alpha_V|^2 = |\alpha|^2 \gg |\alpha_H|^2\approx 0$, resulting in $\hat S_1 =-|\alpha_V|^2$, $\hat S_2^{\prime} =\delta{\hat{X}}^+_{H,\text{vac}}$, and $\hat S_3^{\prime}=\delta{\hat{X}}^-_{H,\text{vac}}$.
Note that the subscript $(\cdot)_{\text{vac}}$ indicates that unmodulated $\hat S_2^{\prime}$ and $\hat S_2^{\prime}$ have zero mean and variance equal to that of vacuum fluctuations.

\subsection{Quantum encoding} \label{Sec: QuantEnc}
After encoding the classical information, Alice then directs the output beam to her QKD modulator. In this work, we follow the existing experimental demonstrations~\cite{Lorenz2004,Lorenz2006,heim2010,Stokes2014Discrete,Elser2009} and assume Alice has a QKD modulator consisting of a magneto-optic modulator (MOM), which introduces a controlled weak coupling from $\hat{S}_1$ to $\hat{S}_2$ (while keeping $\hat{S}_3$ unchanged), and an electro-optic modulator (EOM), which introduces a controlled weak coupling from $\hat{S}_1$ to $\hat{S}_3$ (while keeping $\hat{S}_2$ unchanged). This corresponds to Step {\raisebox{.5pt}{\textcircled{\raisebox{-.9pt} {3}}}} in Fig.~\ref{Fig.Schematic}.

For each shot, Alice modulates $\hat S_2$ ($\hat S_3$) with a small random real number via a MOM (EOM), effectively adding a small fluctuation term $\delta{\hat{X}}^+_{M,\text{QKD}}$ ($\delta{\hat{X}}^-_{M,\text{QKD}}$) to the unmodulated $\hat S_2^{\prime}$ ($\hat S_3^{\prime}$) in Eq.~(\ref{Eq:S2_normalised}) (Eq.~(\ref{Eq:S3_normalised})).
According to~\cite{Lorenz2004}, by controlling the modulation current through the MOM coil (modulation voltage applied to the EOM), $\delta{\hat{X}}^+_{M,\text{QKD}}$ ($\delta{\hat{X}}^-_{M,\text{QKD}}$) can be continuously adjusted according to the random number Alice draws. 
{{Specifically, Alice's modulation current through her MOM and modulation voltage applied to her EOM are adjusted such that the small QKD modulations $\delta{\hat{X}}^+_{M,\text{QKD}}$ and $\delta{\hat{X}}^-_{M,\text{QKD}}$ are 
independent and follow a Gaussian distribution with zero mean and variance $V_{\text{mod}}$.}}
After the quantum modulation, the modulated $\hat S_2^{\prime}$ and $\hat S_3^{\prime}$ becomes
\begin{equation}\label{Eq:S2Modulated}
    \hat S_2^{\prime}= \delta{\hat{X}}^+_{M} = \delta{\hat{X}}^+_{M, \text{vac}}+\delta{\hat{X}}^+_{M,\text{QKD}},
\end{equation}
 \begin{equation}\label{Eq.S3Modulated}   
    \hat S_3^{\prime}=\delta{\hat{X}}^-_{M} = \delta{\hat{X}}^-_{M, \text{vac}}+\delta{\hat{X}}^-_{M,\text{QKD}}.
\end{equation}
The variance of the modulated $\hat S_2^{\prime}$ and $\hat S_3^{\prime}$ then become
\begin{align} \label{Eq.Variance}
    \operatorname{Var}(\hat S_2^{\prime})= \operatorname{Var}(\hat S_3^{\prime})=1+V_{\text{mod}}.
\end{align}
}

\subsection{Transmission} \label{Sec: Transmit}
After performing the classical-quantum modulation, Alice transmits her beam to Bob through an FSO channel with transmissivity $T$, given as $T=10^\frac{-\mathcal{L}}{10}$, where $\mathcal{L}$ is the channel loss in dB. This step corresponds to Step {\raisebox{.5pt}{\textcircled{\raisebox{-.9pt} {4}}}} in Fig.~(\ref{Fig.Schematic}, \ref{Fig:AliceState}). We assume that Alice is a satellite transmitting to a ground receiver, Bob, and, as shown by our own numerical phase-screen simulations~\cite{SQCC_IOP}, approximate such a downlink channel as diffraction-dominated (fixed $T$) - determined by the sizes of the transceiver apertures.

\subsection{Classical decoding}  \label{Sec: ClassDeco}
Bob's classical decoding corresponds to Step {\raisebox{.5pt}{\textcircled{\raisebox{-.9pt} {5}}}} in Fig.~\ref{Fig.Schematic}.
When Bob receives Alice's signal, he first divides it into transmitted and reflected components using a beam splitter (BS1) with transmittance $\eta$. In order to determine the encoded classical bit, Bob directs the transmitted component of BS1 into a decoding module, which splits its input using a polarizing beam splitter (PBS), performs a direct detection on each of the two PBS outputs using a photodetector (e.g., photodiode), and then takes the difference between the measurement results.
This operation effectively measures $\hat S_1 = \hat n_H - \hat n_V$ in Eq.~(\ref{Eq:S1}). If the measured result of $\hat S_1$ is positive, Bob would decode the classical bit as ``1'' (recall Eq.~(\ref{Eq:S1_AC})), and if the measured result of $\hat S_1$ is negative, he will decode the classical bit as ``0''. Bob further records his measurement result of $\hat S_1$ for the later steps.  

We use the bit error rate (BER) as our performance metric for classical communications -- a bit error occurs when Alice transmits the classical bit ``0'' (``1'') and Bob gets ``1'' (``0'') from his classical decoding. There are two sources of noise that affect the faithful transfer of the classical bit, namely vacuum noise and electronic noise $\nu_{\text{el}}$. 
The BER for the classical communication part of our protocol  is given by~\cite{ber2,CBER}
\begin{equation}\label{Eq:BER}
C_{\text{BER}}= \frac{1}{2} \text{erfc}\left( \frac{\sqrt{2T\eta}\alpha}{\sqrt{(1+\nu_{\text{el}})}}\right),
\end{equation}
where $\alpha=\sqrt{|\langle \hat{S}_1\rangle|}$, and $\operatorname{erfc}(\cdot)$ denotes the complementary error function.

It is worth pointing out that unlike the conventional SQCC protocol (where the QKD modulation imposes a noise-like effect on the decoding of classical information{\color{red}~\cite{Nick2025}}), in our protocol the Gaussian modulations of $\hat S_2$ and $\hat S_3$ (used for QKD encoding) do not contribute to any noise in the decoding of the classical bit -- this is due to our careful design of the classical-quantum encoding. 
Our classical encoding ensures that the classical amplitudes $\alpha_H$ and $\alpha_V$ cannot be nonzero at the same time, and our quantum encoding ensures that the zero-mean quantum modulation is always applied to the weakly polarized component whose classical amplitude is zero. This design ensures that both $\langle \hat{S}_1\rangle$ and $\operatorname{Var}(\hat{S}_1)$ do not depend on quantum modulation, eliminating any potential interference imposed by quantum modulation on the decoding of the classical bit. 

\subsection{Quantum decoding}  \label{Sec:QuantDeco}
After classical decoding, Bob proceeds to his quantum decoding, which involves measuring $\hat S_2$ and $\hat S_3$ followed by a normalization. As illustrated in Step {\raisebox{.5pt}{\textcircled{\raisebox{-.9pt} {6}}}} of Fig.~\ref{Fig.Schematic}, Bob uses an optical switch that randomly directs the reflected component of BS1 (recall Section~\ref{Sec: ClassDeco}) towards his setup for the measurement of $\hat S_2$ or $\hat S_3$. Just like in the classical decoding, Bob's quantum decoding is solely based on the use of direct detection via photodiodes.

Specifically, in the measurement setup of $\hat S_2$, the reflected component of BS1 is directed through a half-wave plate (HWP) positioned at $22.5^{\circ}$, which is the angle between the fast axis of HWP and plane of polarization of the beam (this convention holds for all wave plate angles mentioned hereafter), followed by a decoding module to measure the difference between the number of photons in the $D$ (diagonal) component (i.e., $\hat n_D$) and the $A$ (anti-diagonal) component (i.e., $\hat n_A$).  
Similarly, in the measurement setup of $\hat S_3$, the reflected component of BS1 is directed through an HWP at an angle of $22.5^{\circ}$, a quarter-wave plate (QWP) at $45^{\circ}$ ,  followed by a decoding module to determine the difference between the numbers of photons in the $R$ (right-handed circular) component ($\hat n_R$) and the $L$ (left-handed circular) component ($\hat n_L$). 

Bob further normalizes his measurement results of $\hat{S}_2$ and $\hat{S}_3$ by his recorded measured value of $\hat S_1$ (with channel transmissivity $T$ taken into account) by classical post-processing (as shown in Step {\raisebox{.5pt}{\textcircled{\raisebox{-.9pt} {7}}}} of Fig.~\ref{Fig.Schematic}.) In doing so, Bob effectively extracted the values of $\hat S_2^{\prime}$ and $\hat S_3^{\prime}$. After repeating the Steps {\raisebox{.5pt}{\textcircled{\raisebox{-.9pt} {1}}}} to {\raisebox{.5pt}{\textcircled{\raisebox{-.9pt} {7}}}} for a sufficient number of times, Alice and Bob begin the required post-processing, including sifting, parameter estimation, information reconciliation (note that we assume reverse reconciliation) and privacy amplification, eventually producing a final secret key on which Eve has no knowledge.

\subsection{Alternative method of state preparation via Pockels cell} \label{Sec: InitialStatePrep2}
\begin{figure}[h!] 
\centering
\includegraphics[width=1.0\linewidth]{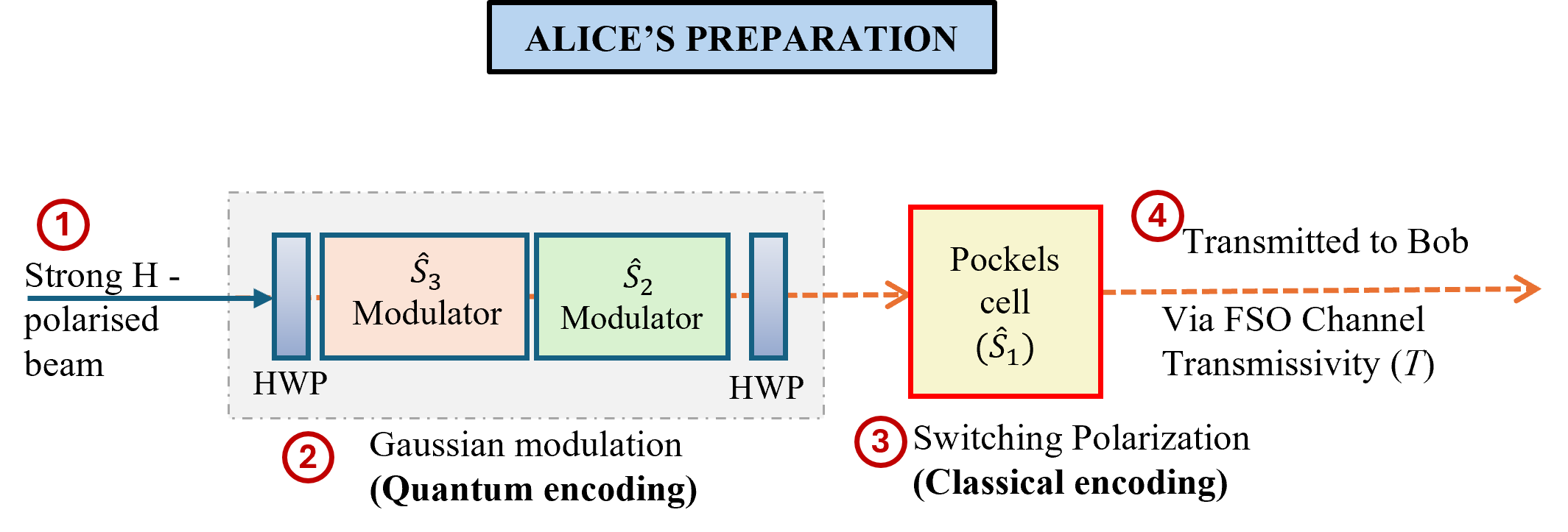}
  \caption{ Schematic diagram of Alice's state preparation method.
   {\raisebox{.5pt}{\textcircled{\raisebox{-.9pt} {1}}}} A strong $H$-polarized beam is directed toward the quantum modulation setup.
   {\textcircled{\raisebox{-.9pt} {2}}}
   The beam passes through HWP set at $22.5^o$, followed by the modulation of $\hat S_3$ and $\hat S_2$ (not necessarily in that order), and another HWP (also set at $22.5^o$). 
   {\raisebox{.5pt}{\textcircled{\raisebox{-.9pt} {3}}}} The modulated beam passes through a Pockels cell for classical encoding. To encode the classical bit as ``0'' (``1''), the Pockels cell switches the beam's polarization to the $V$($H$) direction, respectively.
   {\raisebox{.5pt} {\textcircled{\raisebox{-.9pt} {4}}}} The beam is sent to Bob via an FSO channel with transmissivity $T$.
   Steps {\raisebox{.5pt}{\textcircled{\raisebox{-.9pt} {5}}}} through {\raisebox{.5pt}{\textcircled{\raisebox{-.9pt} {7}}}} remain as shown in Fig.~\ref{Fig.Schematic}, except that in Step {\raisebox{.5pt}{\textcircled{\raisebox{-.9pt} {6}}}} the optical switch (OS) is replaced by a 50:50 beam splitter (BS2). This modification enhances the protocol's practicality by enabling the simultaneous measurement of $\hat S_2$ and $\hat S_3$ (equivalent to heterodyne detection). Note that the setup described in this figure can be directly inserted into Fig.~\ref{Fig.Schematic} to replace the setup before Step {\raisebox{.5pt}{\textcircled{\raisebox{-.9pt} {5}}}} of Fig.~\ref{Fig.Schematic}.}
\label{Fig:AliceState}
\end{figure}

Despite its intuitiveness, the state preparation scheme described in Section~\ref{Sec: InitialStatePrep} is neither efficient nor suitable for practical high-speed operation since it requires the preparation of two beams (in different polarizations) and their controlled combining for classical encoding. Here, we propose a practical and efficient alternative implementation (described in Fig.~\ref{Fig:AliceState}), which is completely equivalent to our proof-of-concept example for the purpose of Stokes-based quantum-classical encoding. 
Specifically, rather than relying on the controlled combining of two orthogonally polarized beams, this alternative implementation only uses a single beam with a fixed polarization and a Pockels cell (which rotates the polarization plane of linearly polarized light by $\pi/2$ when activated) for classical encoding (to be detailed in Section~\ref{Sec: MathForm}). 
In order to accommodate such an improvement, this implementation (illustrated in Fig.~\ref{Fig:AliceState}) performs quantum encoding before classical encoding. 
Note that the setup described in Fig.~\ref{Fig:AliceState} can be directly inserted into Fig.~\ref{Fig.Schematic} to replace the setup before Step {\raisebox{.5pt}{\textcircled{\raisebox{-.9pt} {5}}}} of Fig.~\ref{Fig.Schematic}. In the following, we describe in detail our Stokes-based SQCC protocol based on this alternative implementation.

\begin{figure*}[bt!] \label{Poincare_Sphere}
     \centering
     \includegraphics[width=1 \linewidth]{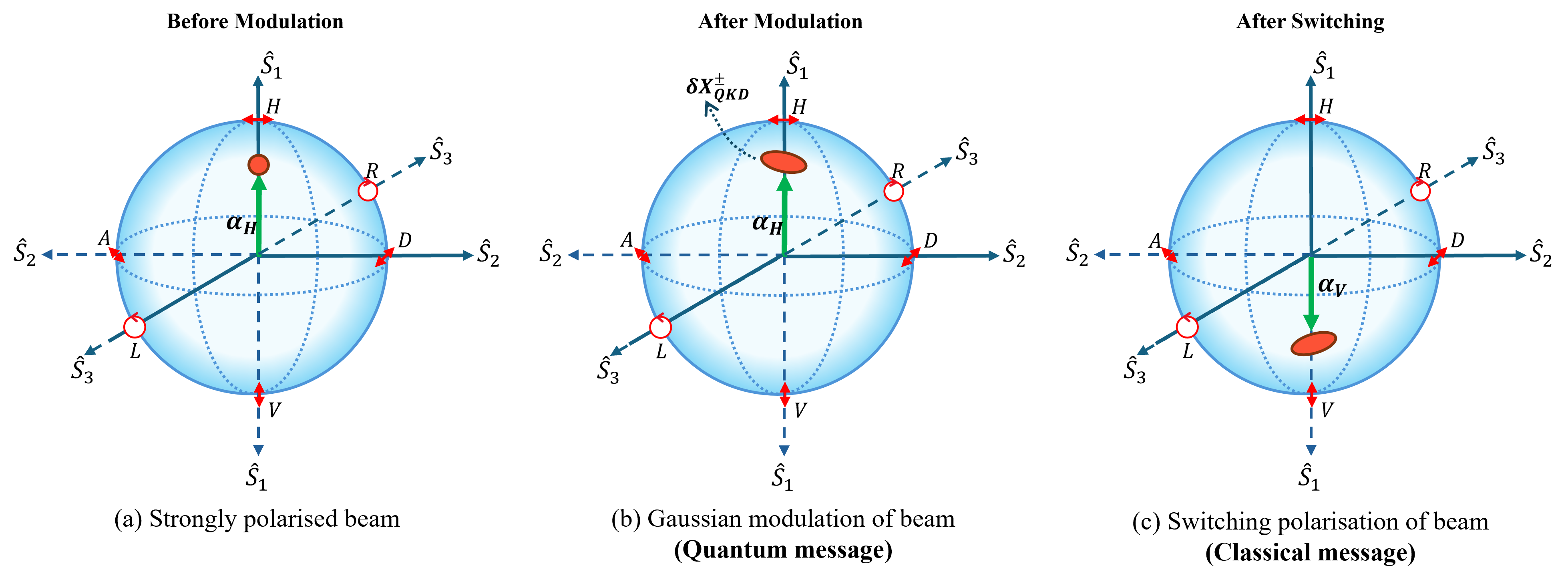}
     \caption{Illustration of the modulation process using Poincare sphere. (a)~Before modulation: the beam is strongly polarized along the $\hat S_1$ axis ($H$-polarization). (b)~After modulation: a small Gaussian perturbation $\delta\hat X^{\pm}_{\text{QKD}}$  is introduced along $\hat S_2$ and $\hat S_3$, encoding the quantum signal. (c)~After switching: the polarization is switched ($H$ to $V$ or vice versa) using a Pockels cell to encode the classical bit, while the quantum signal remains in the weakly polarized component.}
     \label{fig:PoincareSphere}
 \end{figure*}

For each shot of classical-quantum modulation, Alice prepares a single classical beam completely polarized in the $H$ direction with an initial classical amplitude $\alpha$, achieving the exact expressions of Stokes operators (recall Eqs.~(\ref{Eq:S0_AC}-\ref{Eq:S3_AC})) and normalized Stokes operators (recall Eqs.~(\ref{Eq:S2_normalised}-\ref{Eq:S3_normalised})) as if the classical bit ``1'' is encoded. This step corresponds to Step {\raisebox{.5pt}{\textcircled{\raisebox{-.9pt} {1}}}} in Fig.~\ref{Fig:AliceState}.
It is worth pointing out that all the underlying assumptions in Eqs.~(\ref{Eq:S0_AC}-\ref{Eq:S3_normalised}) are still valid since the beam is now completely polarized along the $H$ direction with a negligible $V$ component.
The unmodulated beam after preparation for the initial state is visualized in Fig.~\ref{fig:PoincareSphere}(a) using Poincar\'e sphere representation.

\section{Mathematical formalism} \label{Sec: MathForm}

In this section, we choose to focus on the mathematical framework of the setup described in Section ~\ref{Sec: InitialStatePrep2}, that is, through the configuration of Fig.~\ref{Fig:AliceState}, where the first modulator is an EOM and the second an MOM. \footnote{As seen in Fig.~\ref{Fig:AliceState}, we have generalized the encoding of
$\hat S_2$ and $\hat S_3$ into two ``encoding modules'' that do not necessarily need to be an MOM and an EOM (the two HWPs can also be removed/replaced in alternate setups). This generalization highlights the fact that there are many pathways to achieving our required quantum encoding - a strength of our new protocol.
An example of such an alternate framework is given in \cite{sagnac}, where Sagnac interferometers, phase modulators, beam splitters, and some other simple ancillary equipment are used and an encoding of $\hat S_2$ and $\hat S_3$ suitable for CV-QKD is derived via the Mueller matrix \cite{sagnac}. This leads to a different quantum encoding for the Stokes parameters (see \cite{sagnac} for details).} 
In general, any optical element affecting the Stokes vector can be described mathematically via the Mueller calculus. In this framework, we can write ${{\rm{\textbf{S}}}^{out}} = {\rm{\textbf{M}}}{{\rm{\textbf{S}}}^{in}}$, where {\textbf{M}} is the Mueller 4x4 matrix corresponding to the specific optical element under study (the evolution of the Stokes vector through multiple optical elements occurs through a series of matrix multiplications). The elements $m_{ij}$ of the Mueller matrix can be derived from the Jones matrix, {\textbf{J} (which describes the evolution of the two-dimensional coherent subspace), through ${m_{ij}} = \frac{1}{2}{\rm{Tr}}\left\{ {\textbf{J}{\sigma _i}{\textbf{J}^\dag }{\sigma _j}} \right\}$, where the $\sigma$'s represent the usual Pauli matrices (see \cite{Goldstein}). We determine the Mueller matrix for each of the four components shown in the gray box in Fig.~\ref{Fig:AliceState}, multiply them in sequence to derive the overall matrix for all combined components, and then apply that overall matrix to an input horizontally polarized beam (Stokes vector =[1,1,0,0]$^T$). This leads to the output Stokes vector (see Appendix~B) ${{\rm{\textbf{S}}}^{out}}=[1, \cos {\phi_1} \cos {\phi_2}, \cos {\phi_1} \sin {\phi_2},  \sin {\phi_1}]^T$, where $\phi_1$ is the phase difference introduced by the EOM between the linear polarized basis component (H and V) and
$\phi_2$ is the phase difference introduced by the MOM between the circular polarized basis components (left and right). Taking the small-angle approximation leads to ${{\rm{\textbf{S}}}^{out}}\approx[1, 1, \phi_2,  {\phi_1}]^T$.


 Although the Mueller matrix formalism is ideal for calculations, it is perhaps useful to describe again the changes in the Stokes parameters using a more intuitive pathway. We will not specify the specific nature of the modulation devices so as to keep a general view point. We will consider phase and amplitude modulations to the creation operators,
and these modulations will be driven by an oscillating sine-wave input at radio frequencies, highlighting the importance of sideband separation from the carrier frequency (noise reduction in signal acquisition). \footnote{If the modulators are in fact an EOM followed by an MOM, the sine inputs will correspond to the voltage waveforms driving the modulation in the EOM and the current variations driving the modulations in the MOM.} In our setup
Alice begins with her single beam polarized in the $H$-direction characterized by the classical amplitude $\alpha$, while the $V$-direction is in a vacuum state. Consequently, the total number of photons ($\alpha^2$) is present solely in the $H$ component, and none in the $V$ component.
 Alice then directs the completely $H$ polarized beam into the quantum modulation setup, which consists of two HWPs, and an $\hat{S}_3$ and $\hat{S}_2$ modulators to introduce weak coupling of the Stokes parameter $\hat{S}_1$ to $\hat{S}_3$ and $\hat{S}_2$, respectively. The HWPs are set at an angle of $22.5^\circ$, and the modulators operate at the same driving frequency $\omega_d$. The first HWP, placed before the $\hat{S}_3$ modulator, transforms the completely $H$-polarized beam into a diagonally polarized beam, equally distributing the energy between the $H$ and $V$ components  
\begin{align}
    \langle\hat{a}_H \rangle = \frac{\alpha}{\sqrt2}, \ \langle\hat{a}_V \rangle = \frac{\alpha}{\sqrt2}.
\end{align}
 The $\hat{S}_3$ modulator then introduces a time‐dependent, small phase modulation of the form $a \sin(\omega_d t) $, where $ a $ is the small modulation factor, which remains constant for a single shot. In weak modulation limit, the modulation factor is extremely small $(a \ll 1)$, thus a first-order Taylor expansion can be used to approximate the phase shift $\phi$ between the components
\begin{equation}
    e^{\pm ia\sin(\omega_dt)} \approx 1 \pm ia\sin(\omega_dt).
\end{equation}
This modulation is imposed equally and in opposite directions to the $H$ and $V$ components of the beam, resulting in
\begin{equation}
 \langle\hat{a}_H (t) \rangle  \approx \frac{\alpha}{\sqrt{2}}\left( 1 + i a \sin(\omega_d t) \right),
\end{equation}
\begin{equation}
\langle\hat{a}_V (t)\rangle  \approx \frac{\alpha}{\sqrt{2}}\left( 1 - i a \sin(\omega_d t) \right),
\end{equation}
The beam then passes through the $\hat{S}_2$ modulator. In general, $\hat{S}_3$ modulator devices range from simple Faraday rotators to more complex configurations, which allows for a wide range of modulations such as intensity and phase modulations. Beyond the direction of magnetic field, cavity effects, reflection effects, and interferometric effects can be included, leading to many possible manipulations of the Stokes vector~\cite{mom1,mom2,mom3,mom4}. Here, we assume a configuration of the $\hat{S}_2$ modulator that only applies a time-dependent small-amplitude modulation $ b \cos(\omega_d t) $, introducing further weak coupling between the $H$ and $V$ components, where $b$ is the small-modulation index $(b \ll 1)$ that is constant for a single shot. After first-order approximation, the $H$ and $V$ components can be expressed as
\begin{align}
 \langle\hat{a}_H (t)\rangle & \approx \frac{\alpha}{\sqrt{2}} \left[ 1 - i a \sin(\omega_d t) + b \cos(\omega_d t) \right], \\
 \langle\hat{a}_V (t)\rangle & \approx\frac{\alpha}{\sqrt{2}} \left[  1 + i a \sin(\omega_d t) - b \cos(\omega_d t) \right].
\end{align}
Note that $a$ and $b$ are drawn from independent Gaussian distributions shot by shot. Now, after passing through the modulators, the beam goes through the second HWP set at $22.5^o$, restoring the strong $H$ component, with a small modulation introduced in the $V$ component, which is expressed as
\begin{gather}
  \langle\hat{a}_H (t) \rangle \approx \alpha , \label{Eq:Hmod}  \\
  \langle\hat{a}_V (t) \rangle \approx \alpha  \left[  - i a \sin(\omega_d t) +b \cos(\omega_d t) \right]. \label{Eq:Vmod}
\end{gather}
The full-mode operators are then
\begin{gather}
   \hat a_{0H} = \langle\hat{a}_H(t) \rangle + \delta \hat a_H  \label{Eq:QHmod}, \\
  \hat a_{0V} = \langle\hat{a}_V(t) \rangle + \delta \hat a_V \label{Eq:QVmod}.
\end{gather} } 
This step corresponds to Step {\raisebox{.5pt}{\textcircled{\raisebox{-.9pt} {2}}}} in Fig.~\ref{Fig:AliceState}. 
This process can also be interpreted as Alice effectively introducing small modulation terms, $\delta{{X}}^+_{\text{QKD}}$ and $\delta{{X}}^-_{\text{QKD}}$ to the unmodulated $\hat S_2^{\prime}$ and $\hat S_3^{\prime}$ respectively in Eq.~(\ref{Eq:S2Modulated}, \ref{Eq.S3Modulated}) for the beam, where {$\delta{{X}}^+_{\text{QKD}}= 2 \alpha b\cos(\omega_d t) $ and $\delta{{X}}^-_{\text{QKD}}= 2 i  \alpha a \sin(\omega_d t) $.}
The visual representation of the modulated beam is shown in Fig.~\ref{fig:PoincareSphere}.
{In the set-up of our protocol, the EOM and MOM sinusoidally modulate the optical field over a single shot to allocate power to the sidebands, whose amplitudes are further modulated on a shot by shot basis for QKD modulation. After the mixing down at Bob (to be described in detail later in this section), $\hat S_2^{\prime}$ ($\hat S_3^{\prime}$) becomes equivalent to that described by Eq.~(\ref{Eq:S2Modulated}) (Eq.~(\ref{Eq.S3Modulated})).

As seen in Eq.~(\ref{Eq:Hmod}-\ref{Eq:Vmod}) the modulations imposed encode the quantum signal in the $V$ component, while the amplitude of the $H$ component is only negligibly reduced, ensuring that the optical power of the $H$ component remains effectively constant. This essentially means that the small modulation amplitudes $a$ and $b$ introduced in the component $V$ are so minimal that our initial assumption $|\alpha_H|^2 \gg |\alpha_V|^2\approx 0$ continues to hold. 
The strongly polarized beam after quantum encoding is visualized in Fig.~\ref{fig:PoincareSphere}(b) using Poincar\'e sphere representation.

Alice then proceeds to perform the classical encoding by passing the quantum encoded beam to a Pockels cell, which, once activated, rotates the polarization plane of linearly polarized light by $\pi/2$ so that the originally $H$-polarized component becomes $V$-polarized and vice versa.
The classical bit can be readily encoded using this setup -- Alice encodes bit `0' (`1') by activating (deactivating) the Pockels cell, effectively switching the dominant polarization direction of the quantum-modulated beam depending on the classical bit to be encoded.
This corresponds to Step {\raisebox{.5pt}{\textcircled{\raisebox{-.9pt} {3}}}} in Fig.~\ref{Fig:AliceState}. To encode the classical bit ``0'', Alice switches the polarization of the pulse so that the pulse becomes
\begin{gather}
 \langle\hat{a}_H (t)\rangle \approx\alpha  \left[  - i a \sin(\omega_d t) +b \cos(\omega_d t)\right],\\
  \langle\hat{a}_V (t)\rangle \approx \alpha . \label{Eq:Switching}
\end{gather}
If she desires to send the classical bit ``1'', she leaves the polarization unchanged as in Eq.~(\ref{Eq:Hmod}-\ref{Eq:Vmod}). This simply means that the strongly polarized component of the beam carries the classical bit information, while the weakly polarized component carries the quantum information as visually represented in Fig.~\ref{fig:PoincareSphere}. After this step, the assumption $|\alpha_H|^2 = |\alpha|^2 \gg |\alpha_V|^2\approx 0$ ($|\alpha_V|^2= |\alpha|^2  \gg |\alpha_H|^2\approx 0$) remains valid, which is crucial for our protocol to employ Stokes parameters for simultaneous classical-quantum signaling. The strongly polarized beam after classical decoding is visualized in Fig.~\ref{fig:PoincareSphere}(c) using Poincar\'e sphere representation.



Following the transmission and classical decoding step (as described in Section~\ref{Sec: Transmit}, ~\ref{Sec: ClassDeco}), Bob performs quantum decoding by measuring $\hat S_2$ and $\hat S_3$ followed by normalization. Note that in Fig.~\ref{Fig.Schematic}, Step {\raisebox{.5pt}{\textcircled{\raisebox{-.9pt} 6}}} shows Bob using an optical switch (OS) to detect $\hat S_2$ or $\hat S_3$ (equivalent to the detection of homodyne). However, here we discuss a more practical approach in which Bob employs an additional beam splitter (BS2) (not shown in the figure) to direct the reflected component of BS1  towards his setup for the simultaneous measurement of $\hat S_2$ and $\hat S_3$ (equivalent to heterodyne detection). The reflected component of BS2 is directed towards the $\hat{S}_2$ setup and the transmitted component is directed towards the $\hat{S}_3$ setup.

Now considering the measurement setup of $\hat S_2$, the reflected component of BS2 passes through a half-wave plate (HWP) set at $22.5^{\circ}$ which transforms the linearly polarized beam\footnote{Recall that when Bob receives the beam, it remains a strong linearly polarized, with either strong $H$-component and weak $V$-component or vice-versa.} into the diagonally polarized beam  such that 
\begin{align}
 \langle\hat{a}_H(t) \rangle &= \frac{\sqrt{T}}{2} \frac{\alpha}{\sqrt{2}} \left[ 1 - i a \sin(\omega_d t) + b \cos(\omega_d t) \right], \label{Eq.S2 H-decode} \\
 \langle\hat{a}_V(t) \rangle&= \frac{\sqrt{T}}{2} \frac{\alpha}{\sqrt{2}}  \left[ 1 + i a \sin(\omega_d t) - b \cos(\omega_d t) \right]. \label{Eq.S2 V-decode}
\end{align}
This diagonal beam is then passed through a decoding module, which consists of a polarizing beam splitter (PBS) and photo-diodes to measure the difference between the number of photons in the reflected and transmitted component, which results in  
\begin{align}
    \langle S_{2}(t)\rangle= \frac{T}{4} 2\alpha^2 b \cos(\omega_d t). \label{Eq. S2Decoded}
\end{align}
Similarly, in the $\hat S_3$ measurement setup, the transmitted component from BS2 is routed through a HWP set at $22.5^{\circ}$, transforming the beam into a diagonally polarized (same as Eq.~(\ref{Eq.S2 H-decode}-\ref{Eq.S2 V-decode})). The beam then passes through a quarter-wave plate (QWP) at $45^{\circ}$, which converts it into a circularly polarized beam with its components expressed as 
\begin{align}
    \langle\hat{a}_H(t) \rangle &=\frac{\sqrt{T}}{2} \frac{\alpha}{\sqrt{2}} \left[1  + a \sin(\omega_d t) +  ib \cos(\omega_d t) \right], \label{Eq.S3 H-decode} \\
    \langle\hat{a}_V(t) \rangle &=\frac{\sqrt{T}}{2} \frac{\alpha}{\sqrt{2}}  \left[ 1 -  a \sin(\omega_d t) - ib \cos(\omega_d t) \right]. \label{Eq.S3 V-decode}
\end{align}
After passing through QWP, the beam is directed towards the decoding module (PBS and photodiodes) to determine the difference in photon count between the reflected and transmitted components, which reads as  
\begin{align}
 \langle S_3(t) \rangle = \frac{T}{4} 2 \alpha^2  a \sin(\omega_d t). \label{Eq. S3Decoded}
\end{align}
Noting that the modulations $a \sin(\omega_d t)$ and  $b\cos(\omega_d t)$ we have introduced in the above intuitive analysis correspond to twice the phase differences used in our previous Mueller analysis (i.e., $2\phi_1$ and $2\phi_2$, respectively), we can see a direct correspondence in Eqs.~(\ref{Eq. S2Decoded} and \ref{Eq. S3Decoded}) with our previous determination of ${{\rm{\textbf{S}}}^{out}}\approx[1, 1, \phi_2,  {\phi_1}]^T$. This latter determination is the true representation of the impact on $ \langle S_3 \rangle$ and $\langle S_2 \rangle$, when the $\hat{S}_2$ modulator is an EOM and the $\hat{S}_2$ modulator is an MOM.

In the frequency-domain representation of the decoded signal, the Fourier transforms of Eq.~(\ref{Eq. S2Decoded}, \ref{Eq. S3Decoded}) show the sidebands located at frequencies $\pm \omega_d$. In practical applications, distinct sidebands are maintained by selecting a carrier frequency $(\omega_c)$, which is much higher than the modulating frequency $\omega_d$ (and its bandwidth). This approach prevents the upper and lower sidebands, positioned at $\omega_c \pm  \omega_d$, from overlapping with each other or with adjacent channel sidebands.

So far we have not discussed the time dependence of the classical amplitude. We now make a distinction between $\alpha_t$, the instantaneous amplitude at time $t$ and $\alpha = \bar \alpha_t$, the average amplitude across a signal window. Normalizing by $\sqrt{S_1} = \sqrt{T/2}\alpha_t$ and including the quantum noise, we can now write the normalized operators describing Bob's measurements as:
\begin{align}
    \hat S'_{2}(t) = \sqrt{\frac{T}{2}} \alpha_t b \cos(\omega_d t) + \delta \hat X_t^+, \\
    \hat S'_3(t) = \sqrt{\frac{T}{2}}  \alpha_t  a \sin(\omega_d t) + \delta \hat X_t^-, \label{Eq. S2S3Decoded}
\end{align}
where $\delta \hat X_t^\pm$ represent the instantaneous quantum fluctuations of the amplitude (+) and phase (-) quadratures and includes those coming from Alice and those added during transmission. 
In our system model, Alice prepares identical pulses with a Gaussian temporal profile of width (standard deviation) $\sigma$. Obviously, we need the time window in which each signal is sent to be $> \sigma$. Assuming that the quantum signal modulations are applied with constant amplitude across each pulse, then, in the frequency domain, Eqs. (\ref{Eq. S2Decoded}) and (\ref{Eq. S3Decoded}) will appear as sidebands, centered at $\pm \omega_d$, with widths proportional to $1/\sigma$. Clearly, we need to have $|\omega_d| > 1/\sigma$ to avoid the sidebands overlapping, but we also require that Alice and Bob's systems have low technical noise across the bandwidth of the sidebands so that the signals are quantum limited (as required for QKD).

Given these conditions, Bob can mix-down the sidebands across each detection window to give quantum signals represented by the operators
\begin{align}
    \hat S'_{2}(\omega_d) = \sqrt{\frac{T}{2}} \alpha b + \delta \hat X_{\omega_d}^+, \\
    \hat S'_3(\omega_d) = \sqrt{\frac{T}{2}}  \alpha a + \delta \hat X_{\omega_d}^-. \label{Eq. S2S3Mixed}
\end{align}  
We see that mixing down over a signal window gives us a QKD signal that is proportional to the product of the laser amplitude and the modulation amplitude, plus the quantum noise matched to the side-band modes. Given $a$ and $b$ are picked from independent zero mean Gaussian distributions, we find that averaging over many pulses gives $\alpha \langle a \rangle = \alpha \langle b \rangle = 0$ but $\alpha^2 \langle a^2 \rangle = \alpha^2 \langle b^2 \rangle = V_{\text{mod}}$ (recall that $V_{\text{mod}}$ is the variance of the Gaussian distributions for QKD modulations).
We also have $\langle \delta \hat {X}_{\omega_d}^{\pm} \rangle = 0$ and $\langle {\delta {\hat {X}_{\omega_d}^{\pm}}}^{2} \rangle = T + T \; \Xi_{\text{ch}}$, where the first term on the rhs comes from the quantum noise in Alice's state preparation, whilst the second term is noise introduced by the channel (as will be detailed in the next section).

We now discuss the potential interference from the classical part to the quantum part of communication. A bit error in classical decoding negatively impacts the quantum part of communication via the introduction of additional noise. A bit error occurs in Bob's classical decoding when his measured value of $\hat{S}_1$ differs by a sign from what Alice has prepared and transmitted. Using such a wrong result in his normalization, Bob effectively adds a phase flip of $-\pi/2$ to his extracted values of $\hat S_2^{\prime}$ and $\hat S_3^{\prime}$. For instance, if Alice tries to convey the classical bit ``1'' (by setting $\hat S_1 =|\alpha_H|^2=|\alpha|^2$) but Bob mistakes it for ``0'', Bob will use $-|\alpha_V|^2=-|\alpha|^2$ for his normalization, resulting in a $-\pi/2$ phase between Bob's extracted values of $\hat S_2^{\prime}$ and $\hat S_3^{\prime}$ and their corresponding original values Alice intended to encode for QKD. However, we recall that in deriving Eqs.~(\ref{Eq:S0_1}-\ref{Eq:S3_1}), we have neglected the second-order fluctuation terms. This assumption holds when $\alpha$ at the transmitter and $\alpha’(\sqrt{T\eta}\alpha)$ at the receiver is sufficiently large. To make QKD viable, we estimate that $\alpha '$ must be greater than $10^{2}\ (10^{3})$ at a loss of 10 dB (20 dB) at the receiver, which eventually leads to a classical BER effectively zero (henceforth we set $C_{\text{BER}}=0$.)

We note the conventional SQCC protocol~\cite{Qi2016,Qi2018},  the performance of CV-QKD is heavily limited by phase instability noise, which is proportional to the power allocated to the classical part of communication. Such noise does not exist in our protocol since it does not require the use of a separate LO. 
 We also note that, as long as the aforementioned second-order fluctuation terms can be ignored, our protocol is equivalent to the conventional SQCC protocol~\cite{Qi2018} in terms of security analysis. As a result, the security level of the QKD part of our protocol is the same as~\cite{Qi2018}, which has been argued to be at least as strong as the standard CV-QKD protocol~\cite{R2}. 

 \section{Asymptotic and Finite-size Key Rate Analysis} \label{Sec:Asymp_Finite-sizeKeys}
In practice, pulsed lasers are used more frequently than continuous-wave lasers to perform QKD. Therefore, for our practical performance analysis, we assume that the pulsed laser is quasi-monochromatic (long-duration pulses), ensuring that the mathematical framework described in Section~\ref{Sec: MathForm} remains applicable. We assume that Alice encodes her classical-quantum signal onto identical pulses with a Gaussian temporal envelope such that the total number of photons in the single pulse remains $\alpha^2$. 
The secret key rate is used as a performance metric for the QKD part of our protocol. The noise sources affecting this part include the total channel-added noise $\Xi_{\text{ch}}=\frac{1-T}{T} + \xi$ (where $\xi$ is the total excess noise) and the detector-added noise {$\Xi_{\text{det}}=(2-\eta+2\nu_{\text{el}})/\eta$}.
The total noise affecting quantum communication is given by $\Xi_{\text{tot}}= \Xi_{\text{ch}} + \frac{\Xi_{\text{det}}}{T}$. 


Although the QKD part of our protocol is described as a prepare-and-measure scheme, we follow common practice and use the equivalent entanglement-based description for our performance analysis.
Since $\hat S_2^{\prime}$ ($\hat S_3^{\prime}$) is equivalent to the field quadrature $\hat{Q}$ ($\hat{P}$) in coherent-state CV-QKD, a virtual two-mode-squeezed vacuum (TMSV) state can be established with covariance matrix (CM)
\begin{equation}\label{Eq:CM_Original}
\scalebox{0.9}{%
    $\Sigma_{\text{AB}}= \begin{bmatrix}
        V\mathbb{I}_2 & \sqrt{V^2-1} \sigma_z\\
        \sqrt{V^2-1} \sigma_z & V\mathbb{I}_2
    \end{bmatrix},$}
\end{equation}
where A (B) denotes the index of the first (second) mode, $V=\operatorname{Var}(\hat{S}_2^{\prime})=\operatorname{Var}(\hat{S}_3^{\prime})$ which is given by Eq.~(\ref{Eq.Variance}), $\sigma_z= \operatorname{diag} (1,-1)$, and $\mathbb{I}_2 =\operatorname{diag} (1,1)$. 
In the entanglement-based description, mode B of the TMSV state is sent through the FSO channel from Alice to Bob. 
Taking into account the channel loss and all the aforementioned noise terms, the CM of the resulting two-mode state is given by
\begin{equation}\label{Eq:CM_Output}
\scalebox{0.9}{%
    $ \Sigma_{\text{AB}}^{\prime}= \begin{bmatrix} 
        V\mathbb{I}_2             & \sqrt{ T(V^2-1)}\sigma_z\\
        \sqrt{T(V^2-1)}\sigma_z  & T(V+\Xi_{\text{ch}})\mathbb{I}_2
    \end{bmatrix}.$}
\end{equation}
Note that, we assume that Eve does not have control over $\nu_{\text{el}}$ and $\eta$ since both of them are inside Bob’s system. 
In the asymptotic regime, the secret key rate of the QKD part of our protocol is given by~\cite{R3}
\begin{equation}\label{Eq:SKR}
    r=\beta I_{\text{AB}}-\chi_{\text{EB}},
\end{equation}
where $\beta$ is the reconciliation efficiency, $\chi_{\text{EB}}$ is the Holevo information between Bob and Eve, and $I_{\text{AB}}$ is the mutual information between Alice and Bob, which is given by
\begin{equation}
    I_{\text{AB}}= \log_2\left[ \frac{V+\Xi_{\text{tot}}}{1+\Xi_{\text{tot}}}\right] \ .
\end{equation}
Assuming Eve applies the collective attack (where she applies the same attack to all quantum states and performs an optimal collective measurement on them at any later time), the Holevo information between Bob and Eve is given by \cite{eqcvqkd}
\begin{align}\label{Eq:chiBE}
    \chi_{\text{BE}} = G(\lambda_1)+G(\lambda_2) - G(\lambda_3) - G(\lambda_4),
\end{align}
where $G(x)=\left(\frac{x+1}{2}\right)\log_2 \left(\frac{x+1}{2}\right)-\left(\frac{x-1}{2}\right)\log_2\left(\frac{x-1}{2}\right)$. According to~\cite{Qi2018}, $\lambda_{1,2,3,4}$ in Eq.~(\ref{Eq:chiBE}) are given by
\begin{equation}\label{Eq:SymEig12}
    \lambda_{1,2} = \sqrt {\frac{1}{2}\left[A \pm \sqrt{A^2-4B} \right]},
\end{equation}
\begin{equation}\label{Eq:SymEig34}
\lambda_{3,4}=\sqrt{\frac{1}{2}\left[C \pm \sqrt{C^2-4 D}\right]}.
\end{equation}
In Eq.~(\ref{Eq:SymEig12}), $A$ and $B$ are given by $A = V^2\left(1-2 T\right)+2 T+T^2\left(V+\Xi_{\text {ch}}\right)^2$ and $B = T^2\left(V \Xi_{\text{ch}}+1\right)^2$. In Eq.~(\ref{Eq:SymEig34}), $C$ and $D$ are given by
\begin{align}
C &= \frac{1}{[T (V + \Xi_{\text{tot}})]^2} \Big\{ A (\Xi_{\text{det}})^2 + B + 1 \notag \\
  &\quad + 2 \Xi_{\text{det}} \big[V \sqrt{B} + T (V + \Xi_{\text{ch}})\big] + 2T(V^2 - 1) \Big\}, 
\end{align}
\vspace{-1.5em}
\begin{align}
D &=\left[\frac{V + \sqrt{B} \Xi_{\text{det}}}{T (V + \Xi_{\text{tot}})}\right]^2.
\end{align}
Using Eqs.~(\ref{Eq:SymEig12}-\ref{Eq:SymEig34}), the value of $\chi_{\text{BE}}$ is calculated via Eq.~(\ref{Eq:chiBE}) to obtain the secret key rate expressed in Eq.~(\ref{Eq:SKR}). Note that in this section, we presented the mathematical formalism for heterodyne detection. For the numerical calculations of secret key rates using homodyne detection (corresponding to Fig.~\ref{Fig.Schematic}), we follow the approach of \cite{Qi2016}, and the respective results are provided in the next section.

For the composable finite-size key regime, we consider that Alice sends $N$  coherent states to Bob (or $N$ two-mode squeezed vacuum states in the equivalent entanglement-based scheme). A fraction $m$ of these coherent states is used for parameter estimation and the remaining $n$ coherent states are used for key generation. Following the approach in ~\cite{ImprovedKeys}, we compute Eve’s Holevo bound while considering the worst-case channel transmissivity $(T)$ and excess noise in the channel $(\Xi_\text{ch})$. The security analysis for composable finite-size key rates in~\cite{ImprovedKeys} provides a more refined formulation compared to the previously existing literature ~\cite{Leverrier, Lupo, Papanastasiou, Weedbrook2004,Pirandola2015,papanastasiou2023composable}. The detailed numerical analysis is provided in the next section.

\section{Numerical Analysis}\label{Sec:Results}
In this section, we perform a numerical evaluation on the performance of our Stokes--based SQCC protocol. 
Specifically, following~\cite{Qi2018,ImprovedKeys}, we set $\eta=0.5$, $\xi=0.01$, $\beta = 0.95$, and $\nu_{\text{el}}=0.1$ for the calculations of the key rates of the asymptotic regime. For finite-size key rates, we set all security parameters $\epsilon$ to $2^{-32}$, probability of successful error correction, $p_{ec}$ to $0.95$, and the fraction of states sacrificed for parameter estimation to $m=N/10$.
Since quantum modulation in QKD does not impose any interference on the decoding of the classical bit (recall Section~\ref{Sec: MathForm}), our protocol allows the introduction of CV-QKD without affecting the guaranteed classical QoS.
\begin{figure}[h!] 
    \centering
    \includegraphics[width=1.08\linewidth]{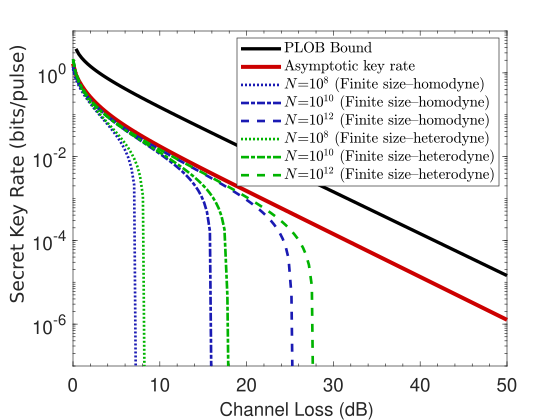}
    \caption{ Secret key rate as a function of channel loss $\mathcal{L}$ for homodyne (blue) and heterodyne (green) detection in the finite-size regime, compared to the asymptotic regime (red) and the PLOB bound (black). The parameters are set as $\eta=0.5$, $\xi=0.01$, $\beta = 0.95$, $\nu_{\text{el}}=0.1$ and the modulation variance $V_\text{mod}$ is numerically optimized at each channel loss.}
    \label{Fig:HeteroHomoCurves}
\end{figure}
In Fig.~\ref{Fig:HeteroHomoCurves}, we present the optimized secret key rate as a function of channel loss (in dB) for homodyne and heterodyne detection schemes in the finite-size regime with various blocks ($N$). The black curve represents the PLOB (Pirandola-Laurenza-Ottaviani-Banchi) bound, which serves as the fundamental upper limit on the secret key rate for any repeaterless QKD protocol \cite{PLOB}. The solid red curve denotes the asymptotic key rate, showing an exponential decrease in key rate with increasing channel loss.
 For finite-size scenarios, the blue and green curves, respectively, depict the secret key rates for homodyne and heterodyne key rates for different block sizes ($N =10^8 $, $10^{10}\  \text{and} \ 10^{12}$). It can be seen that for smaller block sizes the key rate drops significantly and the tolerable channel loss is considerably lower. Comparison of homodyne (blue) and heterodyne (green) detection reveals that homodyne detection typically degrades faster, while heterodyne detection remains robust for slightly higher loss values. For example, the blue dotted-dashed curve representing the finite-size key rates for homodyne detection with block size $N=10^{10}$ begins to decline around $16$ dB of channel loss, while the corresponding green curve for heterodyne detection shows a decrease at approximately $18$ dB of channel loss. However, with larger block sizes, both detection schemes exhibit finite-size key rates that approach the asymptotic regime. 
\section{Conclusion}\label{Sec:Conclusions}
We proposed a Stokes-based SQCC protocol – one that uses polarization encoding via Stokes operators for classical-quantum signaling. We mapped each step of our new protocol to a highly feasible physical implementation, provided a detailed mathematical description, and explained the corresponding important practical implications. We then performed a detailed performance investigation on the protocol over FSO channels, confirming its usefulness and performance through both theoretical analysis and numerical evaluation. Our new design offers superior real-world practicability by eliminating the need for sending a separate LO, allowing for the simultaneous readout of both the quantum and the classical information using a simple direct detection, and minimizing the undesirable interplay between the quantum and classical parts of communications.

 
\appendix
\section*{Appendix A}\label{appA}
Our protocol relies on a first-order approximation, which assumes that second-order fluctuations are negligible for large initial amplitudes $\alpha$ (recall Eqs.~(\ref{Eq:S0_1}-\ref{Eq:S3_1})). 
The first-order approximation allows us to exploit the equivalence between the normalized Stokes operator $\hat{S}_2^\prime$ ($\hat{S}_3^\prime$) and the field quadrature $\hat{Q}$ ($\hat{P}$) , reducing the security analysis of the QKD component of our Stoke-based SQCC protocol to that of the standard coherent state CV-QKD. 
Second-order fluctuations are usually well suppressed at Alice due to the use of a large initial amplitude $\alpha$; however, this may not be necessarily true at Bob's receiver, where the amplitude $\alpha'$ before detection can be significantly smaller than Alice's initial amplitude $\alpha$ due to, e.g., heavy loss. When $\alpha'$ falls below some critical threshold, second-order fluctuations will become non-negligible. In the following, we discuss the impact of second-order fluctuations on our Stokes-based SQCC protocol.

When the second-order fluctuation terms are taken into account, by setting the condition $|\alpha_H|^2 \gg |\alpha_V|^2\approx 0$, the Stokes operators in Eqs.~(\ref{Eq:S0_1}-\ref{Eq:S3_1}) are  expressed as
\begin{align}
\hat{S}_0 &= \hat{S}_1 = |\alpha_H|^2 + \alpha_H\,\delta\hat{X}_H^+ + \delta\hat{a}_H^\dagger\,\delta\hat{a}_H,\label{Eq:S0_2}\\[1ex]
\hat{S}_2 &= \alpha_H\,\delta\hat{X}_V^+ + \delta\hat{a}_H^\dagger\,\delta\hat{a}_V + \delta\hat{a}_V^\dagger\,\delta\hat{a}_H,\label{Eq:S2_2}\\[1ex]
\hat{S}_3 &= \alpha_H\,\delta\hat{X}_V^- + i(\delta\hat{a}_H^\dagger\,\delta\hat{a}_V + \delta\hat{a}_V^\dagger\,\delta\hat{a}_H).\label{Eq:S3_2}
\end{align}
The corresponding variances of the Stokes operators (originally expressed in Eqs.~(\ref{Eq.VS1}-\ref{Eq.VS3})) then become
\begin{equation}
\scalebox{0.8}{%
$\operatorname{Var}(\hat{S}_0)= \operatorname{Var}(\hat{S}_1)= \alpha_H^2 \langle(\delta{\hat{X}}_H^+)^2\rangle +  \langle(\delta \hat{a}_H^\dagger )^2\rangle \langle(\delta \hat{a}_H)^2\rangle,$}
\end{equation}
\begin{equation}
\scalebox{0.8}{%
$\operatorname{Var}(\hat{S}_2)=\alpha_H^2 \langle(\delta{\hat{X}}_V^+)^2\rangle +  \langle(\delta \hat{a}_H^\dagger )^2\rangle \langle(\delta \hat{a}_V)^2\rangle + \langle(\delta \hat{a}_V^\dagger )^2\rangle \langle(\delta \hat{a}_H)^2\rangle,$}
\end{equation}
\begin{equation}
\scalebox{0.8}{%
$\operatorname{Var}(\hat{S}_3) =\alpha_H^2 \langle(\delta{\hat{X}}_V^-)^2\rangle -  \langle(\delta \hat{a}_V^\dagger )^2\rangle \langle(\delta \hat{a}_H)^2\rangle - \langle(\delta \hat{a}_H^\dagger )^2\rangle \langle(\delta \hat{a}_V)^2\rangle. $}
\end{equation}
At the receiver, the effective amplitude is reduced to \(\alpha_H'\). Assuming the use of the quantum-classical modulation described in Section~\ref{Sec: InitialStatePrep2}, the variances of the normalized Stokes operators (originally expressed in Eq.~(\ref{Eq.Variance})) can be re-expressed as
\begin{align} \label{Eq:S1_Second_order}
 \text{Var}(\hat{S}_1) = \langle (\delta \hat{X}_H^+)^2 \rangle +\frac{1}{\alpha'^2},
\end{align} 
\begin{align} \label{Eq:S2_Second_order}
 \text{Var}(\hat{S}_2') \geq \langle (\delta \hat{X}_V^+)^2 \rangle + \frac{2}{\alpha'^2},
\end{align}
\begin{align} \label{Eq:S3_Second_order}
 \text{Var}(\hat{S}_3') \geq \langle (\delta \hat{X}_V^-)^2 \rangle + \frac{2}{\alpha'^2},
\end{align}
where the variance of the second-order fluctuation term $1/\alpha'^2$ is inversely proportional to $\alpha'^2$ (square of the received amplitude $\alpha'$) and will become non-negligible when $\alpha'$ is not sufficiently large, causing the first-order approximation to fail. 
To some extent, the contribution of the second-order fluctuations can be modeled as untrusted excess noise that degrades the QKD performance of our Stokes-based SQCC protocol. 
In the extreme case where $\alpha'$ approaches zero, such excess noise will dominate the variance of the normalized Stokes operators, rendering QKD non-viable.

We further notice that a practical deployment of our protocol will never operate in a regime where the second-order fluctuations are non-negligible, even without the consideration of the second-order fluctuations.  This is simply because our protocol relies on the use of $\hat{S}_1 \sim |\alpha^\prime|^2$ as an effective LO and, under any loss, requires Alice to choose a sufficiently large initial amplitude $\alpha$ that gives a large $\alpha^\prime$ (on the order of $10^{3}$) for effective measurement. Under such a scenario, one can readily see that the large $\alpha^\prime$ not only renders the second-order fluctuations negligible but also drives the classical BER to effectively zero.

\section*{Appendix B} \label{appB}
The Jones matrix for the half waveplate, the EOM and the MOM are
 $${\textbf{J}_{\lambda /2}}\left( \theta  \right) = \left[ {\begin{array}{*{20}{c}}
   {\cos 2\theta } & {\sin 2\theta }  \\
   {\sin 2\theta } & {-\cos 2\theta }  \\
\end{array}} \right]  ,$$

 $${\textbf{J}_{EOM}}\left( {{\phi _1}} \right) = \left[ {\begin{array}{*{20}{c}}
   { 1 } & 0  \\
   0 & \exp \left( { - i{\phi _1}} \right ) \\
\end{array}} \right]  , $$ 
and
 $${\textbf{J}_{MOM}}\left( {{\phi _2}} \right) = \frac{1}{2}\left[ {\begin{array}{*{20}{c}}
   1 & {1}  \\
   i & -i  \\
\end{array}} \right]\left[ {\begin{array}{*{20}{c}}
   {\exp \left( { - i{\phi _2}} \right)} & 0  \\
   0 & 1  \\
\end{array}} \right]\left[ {\begin{array}{*{20}{c}}
   1 & {1}  \\
   1 & -i  \\ \end{array}} \right]^{-1},$$
respectively.
We construct the corresponding Mueller matrices $\textbf{M}_\frac{\lambda}{2}$, etc., by populating the matrix elements based on
${m_{ij}} = \frac{1}{2}{\rm{Tr}}\left\{ {\textbf{J}{\sigma _i}{\textbf{J}^\dag }{\sigma _j}} \right\}$. Here, ${\sigma _1} = \left[ {\begin{array}{*{20}{c}}
   1 & 0  \\
   0 & 1  \\
\end{array}} \right]$,
${\sigma _2} = \left[ {\begin{array}{*{20}{c}}
   1 & 0  \\
   0 & -1  \\
\end{array}} \right]$,
 ${\sigma _3} = \left[ {\begin{array}{*{20}{c}}
   0 & 1  \\
   1 & {0}  \\
\end{array}} \right]$, and 
${\rm{ }}{\sigma _4} = \left[ {\begin{array}{*{20}{c}}
   0 & { - i}  \\
   i & 0  \\
\end{array}} \right]$. We can then determine ${\textbf{S}^{out}}$ as
 $${\textbf{M}_{\lambda /2}}\left( { - \theta } \right){\textbf{M}_{MOM}}\left( {{\phi _2}} \right){\textbf{M}_{EOM}}\left( {{\phi _1}} \right){\textbf{M}_{\lambda /2}}\left(\theta  \right)\left[ {\begin{array}{*{20}{c}}
   1  \\
   1  \\
   0  \\
   0  \\
\end{array}} \right],$$ where $\theta=\pi/8$.



\bibliographystyle{IEEEtran}
\bibliography{IEEEabrv,references}

\begin{thebibliography}{10}
\providecommand{\url}[1]{#1}
\csname url@samestyle\endcsname
\providecommand{\newblock}{\relax}
\providecommand{\bibinfo}[2]{#2}
\providecommand{\BIBentrySTDinterwordspacing}{\spaceskip=0pt\relax}
\providecommand{\BIBentryALTinterwordstretchfactor}{4}
\providecommand{\BIBentryALTinterwordspacing}{\spaceskip=\fontdimen2\font plus
\BIBentryALTinterwordstretchfactor\fontdimen3\font minus \fontdimen4\font\relax}
\providecommand{\BIBforeignlanguage}[2]{{%
\expandafter\ifx\csname l@#1\endcsname\relax
\typeout{** WARNING: IEEEtran.bst: No hyphenation pattern has been}%
\typeout{** loaded for the language `#1'. Using the pattern for}%
\typeout{** the default language instead.}%
\else
\language=\csname l@#1\endcsname
\fi
#2}}
\providecommand{\BIBdecl}{\relax}
\BIBdecl

\bibitem{GCversion}
A.~Dhiman, Z.~Wang, T.~C. Ralph, R.~Aguinaldo, and R.~Malaney, ``Classical-quantum signaling via {S}tokes parameters,'' in \emph{IEEE Global Communications Conference (Globecom), S. Africa}, 2024, pp. 3497--3502.

\bibitem{R1}
C.~Bennett and G.~Brassard, ``Quantum cryptography: Public key distribution and coin tossing,'' \emph{Theoretical Computer Science}, vol. 560, pp. 175--179, 2014.

\bibitem{Shor2000}
P.~W. Shor and J.~Preskill, ``Simple proof of security of the {BB}84 {QKD} protocol,'' \emph{Phys. Rev. Lett.}, vol.~85, pp. 441--444, 2000.

\bibitem{Gisin2002}
N.~Gisin, G.~Ribordy, W.~Tittel, and H.~Zbinden, ``Quantum cryptography,'' \emph{Rev. Mod. Phys.}, vol.~74, pp. 145--195, 2002.

\bibitem{Bedington2017}
R.~Bedington, J.~M. Arrazola, and A.~Ling, ``Progress in satellite {QKD},'' \emph{npj Quantum Information}, vol.~3, pp. 1--13, 2017.

\bibitem{nauerth2013air}
S.~Nauerth, F.~Moll, M.~Rau, C.~Fuchs \emph{et~al.}, ``Air-to-ground quantum communication,'' \emph{Nature Photonics}, vol.~7, pp. 382--386, 2013.

\bibitem{Liao2017}
S.-K. Liao, W.-Q. Cai, W.-Y. Liu, L.~Zhang \emph{et~al.}, ``Satellite-to-ground quantum key distribution,'' \emph{Nature}, vol. 549, pp. 43--47, 2017.

\bibitem{Yin2017}
J.~Yin, Y.~Cao, Y.-H. Li, J.-G. Ren \emph{et~al.}, ``Satellite-to-ground entanglement-based {QKD},'' \emph{Phys. Rev. Lett.}, vol. 119, p. 200501, 2017.

\bibitem{Dequal2020}
D.~Dequal, L.~T. Vidarte, V.~R. Rodriguez, G.~Vallone \emph{et~al.}, ``Feasibility of satellite-to-ground continuous-variable {QKD},'' \emph{npj Quantum Information}, vol.~7, pp. 1--10, 2020.

\bibitem{ZiqingTMQKD}
Z.~Wang, R.~Malaney, and R.~Aguinaldo, ``Temporal modes of light in satellite-to-{E}arth quantum communications,'' \emph{IEEE Commun. Lett.}, vol.~26, pp. 311--315, 2022.

\bibitem{ZiqingOAMQKD}
Z.~Wang, R.~Malaney, and B.~Burnett, ``Satellite-to-{E}arth quantum key distribution via orbital angular momentum,'' \emph{Phys. Rev. Appl.}, vol.~14, p. 064031, 2020.

\bibitem{Tao2021MitigatingTE}
Z.~Tao, Y.~Ren, A.~Abdukirim, S.~Liu, and R.~zhong Rao, ``Mitigating the effect of atmospheric turbulence on orbital angular momentum-based quantum key distribution using real-time adaptive optics with phase unwrapping.'' \emph{Opt. Express}, vol. 29 20, pp. 31\,078--31\,098, 2021.

\bibitem{flamini2018}
F.~Flamini, N.~Spagnolo, and F.~Sciarrino, ``Photonic quantum information processing: a review,'' \emph{Reports on Progress in Physics}, vol.~82, p. 016001, 2018.

\bibitem{Kish2020}
S.~P. Kish, E.~Villase{\~n}or, R.~A. Malaney, K.~A. Mudge, and K.~J. Grant, ``Feasibility assessment for practical continuous variable {QKD} over the satellite-to-earth channel,'' \emph{Quantum Eng.}, vol.~2, 2020.

\bibitem{Laudenbach2017}
F.~Laudenbach, C.~Pacher, C.-H.~F. Fung, A.~Poppe \emph{et~al.}, ``Continuous‐variable {QKD} with {G}aussian modulation - the theory of practical implementations,'' \emph{Advanced Quantum Technologies}, vol.~1, 2017.

\bibitem{R2}
F.~Grosshans and P.~Grangier, ``Continuous variable quantum cryptography using coherent states,'' \emph{Phys. Rev. Lett.}, vol.~88, p. 057902, 2002.

\bibitem{R3}
F.~Grosshans, G.~Assche, J.~Wenger, R.~Brouri, N.~Cerf, and P.~Grangier, ``Quantum key distribution using {G}aussian-modulated coherent states,'' \emph{Nature}, vol. 421, pp. 238--241, 2003.

\bibitem{Jaksch2024}
K.~Jaksch, T.~Dirmeier, Y.~Weiser, S.~Richter \emph{et~al.}, ``Composable free-space continuous-variable {QKD} using discrete modulation,'' \emph{arXiv preprint arXiv:2410.12915}, 2024.

\bibitem{Hajomer2024Experiment}
A.~A. Hajomer, F.~Kanitschar, N.~Jain, M.~Hentschel \emph{et~al.}, ``Experimental composable key distribution using discrete-modulated continuous variable quantum cryptography,'' \emph{arXiv preprint arXiv:2410.13702}, 2024.

\bibitem{Neda2021}
N.~Hosseinidehaj, N.~Walk, and T.~C. Ralph, ``Composable finite-size effects in free-space continuous-variable quantum-key-distribution systems,'' \emph{Phys. Rev. A}, vol. 103, p. 012605, 2021.

\bibitem{Qi2016}
B.~Qi, ``Simultaneous classical communication and {QKD} using continuous variables,'' \emph{Phys. Rev. A}, vol.~94, p. 042340, 2016.

\bibitem{Qi2018}
B.~Qi and C.~C.~W. Lim, ``Noise analysis of simultaneous quantum key distribution and classical communication scheme using a true local oscillator,'' \emph{Phys. Rev. A.}, vol.~9, p. 054008, 2018.

\bibitem{SQCC_25kmFiber}
R.~Kumar, A.~Wonfor, R.~Penty, T.~Spiller, and I.~White, ``Experimental demonstration of single-shot quantum and classical signal transmission on single wavelength optical pulse,'' \emph{Sci. Rep.}, vol.~9, p. 11190, 2019.

\bibitem{LO_Loophole1}
X.-C. Ma, S.-H. Sun, M.-S. Jiang, and L.-M. Liang, ``Local oscillator fluctuation opens a loophole for {E}ve in practical continuous-variable {QKD} systems,'' \emph{Phys. Rev. A}, vol.~88, p. 022339, 2013.

\bibitem{LO_Loophole2}
J.-Z. Huang, S.~Kunz-Jacques, P.~Jouguet, C.~Weedbrook \emph{et~al.}, ``Quantum hacking on quantum key distribution using homodyne detection,'' \emph{Phys. Rev. A}, vol.~89, p. 032304, 2014.

\bibitem{Nick2025}
N.~Zaunders, Z.~Wang, R.~Malaney, R.~Aguinaldo, and T.~C. Ralph, ``Enhanced simultaneous quantum-classical communications under composable security,'' \emph{arXiv preprint arXiv:2505.03145}, 2025.

\bibitem{Vidiella2006}
A.~Vidiella-Barranco and L.~Borelli, ``Continuous variable quantum key distribution using polarized coherent states,'' \emph{International Journal of Modern Physics B}, vol.~20, pp. 1287--1296, 05 2006.

\bibitem{R23}
N.~Korolkova, G.~Leuchs, R.~Loudon, T.~C. Ralph, and C.~Silberhorn, ``Polarization squeezing and continuous-variable polarization entanglement,'' \emph{Phys. Rev. A}, vol.~65, p. 052306, 2002.

\bibitem{schnabel2003stokes}
R.~Schnabel, W.~P. Bowen, N.~Treps, T.~C. Ralph, H.-A. Bachor, and P.~K. Lam, ``Stokes-operator-squeezed continuous-variable polarization states,'' \emph{Phys. Rev. A}, vol.~67, p. 012316, 2003.

\bibitem{Lorenz2004}
S.~Lorenz, N.~Korolkova, and G.~Leuchs, ``Continuous-variable quantum key distribution using polarization encoding and post selection,'' \emph{Appl. Phys. B}, vol.~79, pp. 273--277, 2004.

\bibitem{Lorenz2006}
S.~Lorenz, J.~Rigas, M.~Heid, U.~L. Andersen, N.~L\"utkenhaus, and G.~Leuchs, ``Witnessing effective entanglement in a continuous variable prepare-and-measure setup and application to a quantum key distribution scheme using postselection,'' \emph{Phys. Rev. A}, vol.~74, p. 042326, 2006.

\bibitem{Elser2009}
D.~Elser, T.~Bartley, B.~Heim, C.~Wittmann, D.~Sych, and G.~Leuchs, ``Feasibility of free space quantum key distribution with coherent polarization states,'' \emph{New J. Phys.}, vol.~11, p. 045014, 2009.

\bibitem{heim2010}
B.~Heim, D.~Elser, T.~Bartley, M.~Sabuncu \emph{et~al.}, ``Atmospheric channel characteristics for quantum communication with continuous polarization variables,'' \emph{Appl. Phys. B.}, vol.~98, pp. 635--640, 2010.

\bibitem{Stokes2014Discrete}
B.~Heim, C.~Peuntinger, N.~Killoran, I.~Khan \emph{et~al.}, ``Atmospheric continuous-variable quantum communication,'' \emph{New J. Phys.}, vol.~16, p. 113018, 2014.

\bibitem{Shen2019}
S.-Y. Shen, M.-W. Dai, X.-T. Zheng, Q.-Y. Sun, G.-C. Guo, and Z.-F. Han, ``Free-space continuous-variable quantum key distribution of unidimensional {G}aussian modulation using polarized coherent states in an urban environment,'' \emph{Phys. Rev. A}, vol. 100, p. 012325, 2019.

\bibitem{Zheng2023}
Z.~Zheng, Z.~Chen, L.~Huang, X.~Wang, and S.~Yu, ``Performance analysis of quantum key distribution using polarized coherent-states in free-space channel,'' \emph{Chin. Phys. B}, vol.~32, p. 030306, 2023.

\bibitem{ArbitaryCoupler}
M.~Wang, A.~Ribero, Y.~Xing, and W.~Bogaerts, ``Tolerant, broadband tunable $2\times 2$ coupler circuit,'' \emph{Opt. Express}, vol.~28, pp. 5555--5566, 2020.

\bibitem{SQCC_IOP}
M.~S. Winnel, Z.~Wang, R.~Malaney, R.~Aguinaldo, J.~Green, and T.~C. Ralph, ``Classical-quantum dual encoding for laser communications in space,'' \emph{New J. Phys.}, vol.~26, p. 033012, 2024.

\bibitem{ber2}
K.~Tsujino, D.~Fukuda, G.~Fujii, S.~Inoue \emph{et~al.}, ``Quantum receiver beyond the standard quantum limit of coherent optical communication,'' \emph{Phys. Rev. Lett.}, vol. 106, p. 250503, 2011.

\bibitem{CBER}
K.~Kikuchi, ``Fundamentals of coherent optical fiber communications,'' \emph{Journal of Lightwave Technology}, vol.~34, pp. 157--179, 2016.

\bibitem{sagnac}
X.-T. Zheng, Q.-F. Zhang, J.~yu~Han, J.~Ling, G.~can Guo, and Z.-F. Han, ``Experimental realization of free-space continuous-variable {QKD} based on fiber {S}agnac interferometer,'' \emph{Opt. Lett.}, vol.~48, pp. 4837--4840, 2023.

\bibitem{Goldstein}
D.~H. Goldstein, \emph{\BIBforeignlanguage{en}{Polarized Light}}, 3rd~ed.\hskip 1em plus 0.5em minus 0.4em\relax CRC Press, 2011.

\bibitem{mom1}
H.~Majeed, A.~Shaheen, and M.~S. Anwar, ``Complete {S}tokes polarimetry of magneto-optical {F}araday effect in a {T}erbium {G}allium {G}arnet crystal at cryogenic temperatures,'' \emph{Opt. Express}, vol.~21, pp. 25\,148--25\,158, 2013.

\bibitem{mom2}
R.~M.~A. Azzam, ``Stokes-vector and {M}ueller-matrix polarimetry,'' \emph{J. Opt. Soc. Am. A}, vol.~33, pp. 1396--1408, 2016.

\bibitem{mom3}
J.~Cervantes-L, D.~I. Serrano-Garcia, Y.~Otani, and B.~Cense, ``Mueller-matrix modeling and characterization of a dual-crystal electro-optic modulator,'' \emph{Opt. Express}, vol.~24, pp. 24\,213--24\,224, 2016.

\bibitem{mom4}
J.~N. Hilfiker, N.~Hong, and S.~Schoeche, ``Mueller matrix spectroscopic ellipsometry,'' \emph{Advanced Optical Technologies}, vol.~11, pp. 59--91, 2022.

\bibitem{eqcvqkd}
J.~Lodewyck, M.~Bloch, R.~Garc\'{\i}a-Patr\'on, S.~Fossier \emph{et~al.}, ``Quantum key distribution over $25\,\mathrm{km}$ with an all-fiber continuous-variable system,'' \emph{Phys. Rev. A}, vol.~76, p. 042305, 2007.

\bibitem{ImprovedKeys}
S.~Pirandola and P.~Papanastasiou, ``Improved composable key rates for {CV}-{QKD},'' \emph{Phys. Rev. Res.}, vol.~6, p. 023321, 2024.

\bibitem{Leverrier}
A.~Leverrier, ``Composable security proof for continuous-variable quantum key distribution with coherent states,'' \emph{Phys. Rev. Lett.}, vol. 114, p. 070501, 2015.

\bibitem{Lupo}
C.~Lupo, C.~Ottaviani, P.~Papanastasiou, and S.~Pirandola, ``Continuous-variable measurement-device-independent {QKD}: composable security against coherent attacks,'' \emph{Phys. Rev. A}, vol.~97, p. 052327, 2018.

\bibitem{Papanastasiou}
P.~Papanastasiou and S.~Pirandola, ``Continuous-variable quantum cryptography with discrete alphabets: composable security under collective {G}aussian attacks,'' \emph{Phys. Rev. Res.}, vol.~3, p. 013047, 2021.

\bibitem{Weedbrook2004}
C.~Weedbrook, A.~M. Lance, W.~P. Bowen, T.~Symul, T.~C. Ralph, and P.~K. Lam, ``Quantum cryptography without switching,'' \emph{Phys. Rev. Lett.}, vol.~93, p. 170504, 2004.

\bibitem{Pirandola2015}
S.~Pirandola, C.~Ottaviani, G.~Spedalieri, C.~Weedbrook \emph{et~al.}, ``High-rate measurement-device-independent quantum cryptography,'' \emph{Nature Photonics}, vol.~9, pp. 397--402, 2015.

\bibitem{papanastasiou2023composable}
P.~Papanastasiou, A.~G. Mountogiannakis, and S.~Pirandola, ``Composable security of {CV-MDI-QKD} with secret key rate and data processing,'' \emph{Scientific reports}, vol.~13, p. 11636, 2023.

\bibitem{PLOB}
S.~Pirandola, R.~Laurenza, C.~Ottaviani, and L.~Banchi, ``Fundamental limits of repeaterless quantum communications,'' \emph{Nature Communications}, vol.~8, p. 15043, 2017.

\end{thebibliography}

\end{document}